\pdfoutput=1
\documentclass[11pt,twoside,a4paper,cmspaper,final,collab]{cms-tdr}
\def\svnVersion{1f29649}\def\svnDate{2022/04/18}\def\cmsCernNoTag{CERN-EP-2021-173}\def\cmsCernDate{\today}\def\cmsMessage{Published in the Journal of High Energy Physics as \href{http://dx.doi.org/10.1007/JHEP03(2022)189}{\doi{10.1007/JHEP03(2022)189}.}}
\begin{document}\cmsNoteHeader{FSQ-13-004}

\newcommand{\ptmin}{\ensuremath{\pt{}_{\text{min}}}\xspace}
\newcommand{\ymax}{\ensuremath{y_{\text{max}}}\xspace}
\newcommand{\ptveto}{\ensuremath{\pt{}_{\text{veto}}}\xspace}
\newcommand{\ptave}{\ensuremath{\overline{p}_\mathrm{T}}\xspace}
\newcommand{\sgincl}{\ensuremath{\sigma^{\text{incl}}}\xspace}
\newcommand{\sgmn}{\ensuremath{\sigma^{\mathrm{MN}}}\xspace}
\newcommand{\sgexcl}{\ensuremath{\sigma^{\text{excl}}}\xspace}
\newcommand{\sgveto}{\ensuremath{\sigma^{\text{excl}}_{\text{veto}}}\xspace}
\newcommand{\Rincl}{\ensuremath{R^{\text{incl}}}\xspace}
\newcommand{\RMN}{\ensuremath{R^{\mathrm{MN}}}\xspace}
\newcommand{\Rinclveto}{\ensuremath{R^{\text{incl}}_{\text{veto}}}\xspace}
\newcommand{\RMNveto}{\ensuremath{R^{\mathrm{MN}}_{\text{veto}}}\xspace}
\newcommand{\mur}{\ensuremath{\mu_{\mathrm{r}}}\xspace}
\newcommand{\muf}{\ensuremath{\mu_{\mathrm{f}}}\xspace}
\newcommand{\PYTHIAeight} {\PYTHIA{8}\xspace}
\newcommand{\HERWIGseven} {\HERWIG{7}\xspace}
\newcommand{\HEJ} {{\textsc{hej}}\xspace}
\newcommand{\HEJARIADNE} {{\textsc{hej+ariadne}}\xspace}
\newcommand{\ARIADNE} {{\textsc{ariadne}}\xspace}
\newcommand{\POWHEGPYTHIAeight} {\POWHEG{+}\PYTHIA{8}\xspace}
\newcommand{\POWHEGHERWIG} {\POWHEG{+}\HERWIG\xspace}
\newcommand{\Dy}{\ensuremath{\Delta y}\xspace}

\providecommand{\cmsLeft}{left\xspace}
\providecommand{\cmsRight}{right\xspace}

\cmsNoteHeader{FSQ-13-004} 
\title{Study of dijet events with large rapidity separation in proton-proton collisions at \texorpdfstring{$\sqrt{s} = 2.76\TeV$}{sqrt(s) = 2.76 TeV}}

\date{\today}

\abstract{The cross sections for inclusive and Mueller--Navelet dijet production are measured as a function of the rapidity separation between the jets in proton-proton collisions at $\sqrt{s} = 2.76\TeV$ for jets with transverse momentum $\pt > 35\GeV$ and rapidity $\abs{y} < 4.7$. Various dijet production cross section ratios are also measured. A veto on additional jets with $\pt > 20\GeV$ is introduced to improve the sensitivity to the effects of the Balitsky--Fadin--Kuraev--Lipatov (BFKL) evolution. The measurement is compared with the predictions of various Monte Carlo models based on leading-order and next-to-leading-order calculations including the Dokshitzer--Gribov--Lipatov--Altarelli--Parisi leading-logarithm (LL) parton shower as well as the LL BFKL resummation.}

\hypersetup{%
pdfauthor={CMS Collaboration},%
pdftitle={Study of dijet events with large rapidity separation between the jets in pp collisions at sqrt(s)=2.76TeV},%
pdfsubject={CMS},%
pdfkeywords={CMS, forward dijets, large rapidity intervals between jets, BFKL}}

\maketitle 

\section{Introduction}
The hard scattering of hadrons with high transverse momentum \pt is well described within the 
perturbative quantum chromodynamics (QCD) formalism at large center-of-mass energies $\sqrt{s}$ in the 
Bjorken limit. In this limit, $\sqrt{s} \to \infty$, while the scaling variable $x \sim \pt / \sqrt{s}$ is kept fixed near unity. In this framework $\pt^2 \approx Q^2$, where $Q^2$ is the square of the four-momentum transfer. Thus, since $x$ is not small, $Q^2$ tends to infinity, requiring the terms with $[\alpS\ln Q^2]^n$ in the calculations to be resummed, where \alpS is the strong coupling. This resummation can be carried out in the collinear factorization framework with the Dokshitzer--Gribov--Lipatov--Altarelli--Parisi (DGLAP) formalism~\cite{Gribov:1972ri, Gribov:1972rt, Lipatov:1974qm, Altarelli:1977zs, Dokshitzer:1977sg}. 

Another important kinematic regime is the Regge--Gribov limit~\cite{Kuraev:1976ge, Kuraev:1977fs, Balitsky:1978ic}: $x \to 0$, finite $\pt \gg \Lambda_\mathrm{QCD}$, and $\sqrt{s} \to \infty$, where $\Lambda_\mathrm{QCD}$ is the QCD scale. This leads to large rapidity intervals between the scattered partons. The standard DGLAP approach 
fails at small $x$ because of the necessity to sum the terms of the 
perturbation series enhanced by powers of $[\alpS\ln(1/x)]^n$. Resummation of the 
leading $[\alpS\ln(1/x)]$ terms is typically performed in the Balitsky--Fadin--Kuraev--Lipatov 
(BFKL) approach~\cite{Kuraev:1976ge, Kuraev:1977fs, Balitsky:1978ic}, which describes what is generally referred to as the perturbative pomeron. In the Regge--Gribov limit, the parton 
scattering subprocesses involve a large number of the partons emitted in a large rapidity interval with comparable \pt. Such 
subprocesses can be described by the BFKL evolution and lead to an increase of the dijet production cross section with increasing rapidity separation between the jets, 
$\Dy = \abs{y_1 - y_2}$, where $y_1$ and $y_2$ are the rapidities of the two jets constituting a dijet. In events that contain at least two jets with \pt above a \ptmin threshold, the most 
forward and the most backward ones are referred to as Mueller--Navelet 
(MN) jets~\cite{Mueller:1986ey}. The leading contribution to the dijet cross section in the BFKL approach comes from the MN jets production. 

In Ref.~\cite{Mueller:1986ey}, the BFKL approach was used to calculate the ratio of the MN dijet production cross section \sgmn to that calculated for the Born subprocess. It has been assumed that because of factorization, the corresponding 
hadronic dijet ratio is equal to the one at the subprocess parton level. This ratio is known as the MN dijet $K$ factor. However, the cross section of the Born subprocess is not measurable, because it is not possible to neglect virtual contributions to the measured cross section. Nevertheless, one can measure the cross section for events with exactly two jets with $\pt > \ptmin$. In the following, we will refer to such events as ``exclusive". The lower the value of \ptmin, the better the ``exclusive" process approximates the truly exclusive case, \ie, when there are only two jets and two protons in the final state. Therefore, the cross section of ``exclusive" events can be an approximate measure of the Born subprocess cross section.

Each pairwise combination of jets with $\pt > \ptmin$ in the event forms an inclusive dijet. Therefore, the MN jet pairs constitute a subset of the inclusive dijets. The inclusive cross section \sgincl may have an advantage with respect to the MN cross section \sgmn because some MN jets can fall outside of the detector acceptance~\cite{Kim:1995zu}. For small \Dy, the inclusive dijet cross section is larger than the corresponding MN one because all jets in the rapidity 
interval between the MN jets contribute to the inclusive cross section. 
For large \Dy, the MN dijet cross section approaches 
that for inclusive dijets because of the kinematical limitations on 
additional jet production. 

The present paper reports a measurement of inclusive and MN dijet differential cross sections as a function of \Dy in proton-proton ($\Pp\Pp$) collisions at $\sqrt{s} = 2.76\TeV$, based on a sample corresponding to an integrated luminosity of 5.4\pbinv collected by the CMS experiment at the CERN LHC in 2013:

\begin{equation}\label{eq:x-sections}
\begin{gathered}
	\ddinline{\sgincl}{\Dy},\\
	\ddinline{\sgmn}{\Dy},	
\end{gathered}
\end{equation}
as well as the following cross section ratios:\begin{equation}\label{eq:k-factors}
\begin{aligned}
	\Rincl =& (\ddinline{\sgincl}{\Dy})/(\ddinline{\sgexcl}{\Dy}),\\
	\RMN =& (\ddinline{\sgmn}{\Dy})/(\ddinline{\sgexcl}{\Dy}),\\
	\Rinclveto =& (\ddinline{\sgincl}{\Dy})/(\ddinline{\sgveto}{\Dy}),\\
	\RMNveto =& (\ddinline{\sgmn}{\Dy}) / (\ddinline{\sgveto}{\Dy}).
\end{aligned}
\end{equation}

Here \sgincl is the inclusive 
dijet production cross section, \ie, the cross section for events with 
at least one pair of jets with $\pt > \ptmin = 35\GeV$. In those events each pairwise combination of jets with $\pt > 35\GeV$ contributes to the inclusive cross section. The ``exclusive" dijet production cross section, \sgexcl, corresponds to dijet events with exactly two jets with $\pt > 35\GeV$. 
The cross section of MN dijets is denoted as \sgmn; here, for each event, only the dijet with
the most forward and most backward jets with $\pt > 35\GeV$ contribute. Finally, \sgveto corresponds to ``exclusive" 
events in which there are no extra jets above $\ptveto = 20\GeV$~\cite{Gavrilov:2013opa}. 
 
 Lowering the veto threshold \ptveto 
makes the \Rinclveto and \RMNveto ratios closer to the theoretical 
$K$~factor and increases the sensitivity to BFKL evolution effects. The event definitions as well as the number of collected dijets are summarized in Table~\ref{tab:event_sel}.

\begin{table*}[htb!]
\centering
\topcaption{Event definitions for cross section measurements.\label{tab:event_sel}}
\renewcommand{\arraystretch}{1.3}
\begin{tabular}{p{2.7cm} >{\centering\arraybackslash}p{2.5cm} >{\centering\arraybackslash}p{2.5cm} >{\centering\arraybackslash}p{4cm} >{\centering\arraybackslash}p{2cm}}

Dijet types & Number of jets with $\pt > 35\GeV$ \newline in the event & Veto on extra jets with $\pt > 20\GeV$ \newline in the event & Dijet selection criteria & Number of dijets \\
\hline
Inclusive & $\geq$2 & no & each pairwise combination of jets & 154\,124 \\
MN & $\geq$2 & no & pair of jets most separated in rapidity & 138\,908 \\
``Exclusive" & $=$2 & no & just one jet pair & 132\,648 \\
``Exclusive" with veto & $=$2 & yes & just one jet pair & 107\,770 \\

\end{tabular}
\end{table*}

Previous searches for BFKL evolution effects in jet production at hadron 
colliders were performed at the Tevatron by the D0 and CDF experiments. D0 
observed a stronger dependence of the MN dijet production cross section on 
the collision energy than expected in the leading-logarithmic (LL) BFKL 
approach~\cite{Abbott:1999ai}. Conversely, no indications of BFKL effects in the MN dijet 
azimuthal decorrelations were observed~\cite{Abachi:1996et}. The D0 results 
on dijet production via hard color-singlet exchange (jet-gap-jet events)~\cite{Abbott:1998jb} are consistent 
with next-to-LL (NLL) BFKL-based calculations~\cite{Ekstedt:2017xxy, Enberg:2001ev}. The CDF experiment observed that dijet production via color-singlet exchange is independent of the
pseudorapidity interval between the jets~\cite{Abe:1997ie}. The CDF results are in
general agreement with BFKL-based calculations.

Both the ATLAS~\cite{Aad:2011jz, Aad:2014pua} and CMS ~\cite{Chatrchyan:2012gwa, Chatrchyan:2012pb, Khachatryan:2016udy} Collaborations at the CERN LHC have measured forward dijet production in $\Pp\Pp$ collisions at $\sqrt{s} = 7\TeV$. The CMS measurements extend 
up to $\Dy = 9.4$ between jets with $\pt > 35\GeV$, whereas the ATLAS measurements extend up to $\Dy = 8$ for dijets with average transverse momentum $\ptave = (\pt{}_1 + \pt{}_2) / 2 > 60\GeV$, where $\pt{}_1$ and $\pt{}_2$ are the transverse momenta of the jets in a dijet 
system. 
The ATLAS measurement of dijet production with no additional jets with $\pt > 20\GeV$ between the jets~\cite{Aad:2011jz} is in agreement~\cite{Aad:2014pua} with the parton-level LL BFKL calculations of the \HEJ Monte Carlo (MC) generator~\cite{Andersen:2011hs}, combined with the dipole parton shower simulation of \ARIADNE~\cite{Lonnblad:1992tz}. The ATLAS measurement of azimuthal decorrelations~\cite{Aad:2014pua} in dijet events is also well described by the combination of \HEJ and \ARIADNE at large \Dy and for $60 <\ptave<200\GeV$. Conversely, in this region, the next-to-leading-order (NLO) parton matrix element prediction of the \POWHEG generator~\cite{Alioli:2010xa}, combined with the LL DGLAP-based parton shower simulated with \PYTHIAeight~\cite{Sjostrand:2007gs} or \HERWIG~\cite{Corcella_2001}, fails to describe the data. The CMS measurement of the cross section for dijet production where one jet is at forward rapidities and one is central is described well by \HEJ~\cite{Chatrchyan:2012gwa}. The CMS results for the dijet cross section ratios~\cite{Chatrchyan:2012pb} are neither reproduced at large \Dy by the LL BFKL-based \HEJARIADNE combination, nor by the LL DGLAP-based \HERWIGpp~\cite{Bahr:2008pv} simulation; they are instead well described by the LL DGLAP-based \PYTHIAeight~\cite{Sjostrand:2007gs} generator. Finally, the azimuthal decorrelations measured by CMS~\cite{Khachatryan:2016udy} are in agreement with NLL BFKL analytic calculations for large \Dy~\cite{Ducloue:2013bva}, whereas the LL DGLAP-based MC generators \PYTHIAeight and \HERWIG cannot describe all the measured observables.
 
In summary, none of the DGLAP-based MC generators is able to reproduce 
all the observables; conversely, NLL BFKL calculations, when available, 
are consistent with the measurements. However, more data at different 
collision energies are necessary, since the energy dependence of the 
DGLAP and BFKL contribution is expected to be different.

The present analysis extends the 7\TeV results~\cite{Chatrchyan:2012pb} by measuring \Rincl, \RMN and, in addition, 	\Rinclveto, \RMNveto and $d\sgincl/d\Dy$, $d\sgmn/d\Dy$ at $\sqrt{s} = 2.76\TeV$. The event selection and jet definition are the same as in the previous measurement, which allows a direct comparison of the results. The integrated luminosity is approximately similar (5.4\pbinv in the present analysis and 5\pbinv at $\sqrt{s} = 7\TeV$). The number of
events decreases because of the lower $\sqrt{s}$. However, the
probability of additional $\Pp\Pp$ interactions within the same or
adjacent bunch crossings, \ie, ``pileup" (PU), is lower in the present
analysis, which reduces the systematic uncertainty. The data were taken 
in special runs with low PU. The average number of collisions per bunch crossing is 
estimated to be 0.35.

The CMS measurements of dijet events with a large
rapidity gap at $\sqrt{s} = 7\TeV$~\cite{Sirunyan:2017rdp} and at $\sqrt{s} = 13\TeV$~\cite{TOTEM:2021rix} are complementary to the work presented here. The
jet-gap-jet analysis excludes events with charged particles in the rapidity gap
between the two leading jets, although allowing them outside the gap. In contrast, the present analysis allows the emission of jets with $\pt < \ptmin$ or \ptveto, both inside and outside the rapidity interval between the two jets. The results in Refs.~\cite{Sirunyan:2017rdp} and ~\cite{TOTEM:2021rix} are in partial agreement with NLL BFKL calculations.

\section{The CMS detector}
The central feature of the CMS apparatus is a superconducting solenoid of 6\unit{m} internal diameter, providing a magnetic field of 3.8\unit{T}. Within the solenoid volume are a silicon pixel and strip tracker, a lead tungstate crystal electromagnetic calorimeter (ECAL), and a brass and scintillator hadron calorimeter (HCAL), each composed of a barrel and two endcap sections. Forward calorimeters extend the pseudorapidity $\eta$ coverage provided by the barrel and endcap detectors. Muons are measured in gas-ionization detectors embedded in the steel flux-return yoke outside the solenoid.

The part of the CMS detector essential to the present analysis is the calorimeter system. The ECAL and HCAL extend to $\abs{\eta} < 3.0$. In the region $\abs{\eta} < 1.74$, the HCAL cells have widths of 0.087 in $\eta$ and 0.087 in azimuth ($\phi$). In the $\eta$-$\phi$ plane, and for $\abs{\eta} < 1.48$, the HCAL cells map on to $5{\times}5$ arrays of ECAL crystals to form calorimeter towers projecting radially outwards from close to the nominal interaction point. For $\abs{\eta} > 1.74$, the coverage of the towers increases progressively to a maximum of 0.174 in $\Delta \eta$ and $\Delta \phi$. Within each tower, the energy deposits in ECAL and HCAL cells are summed to define the calorimeter tower energies, which are subsequently used to provide the energies and directions of hadronic jets. 

The forward hadron (HF) calorimeter uses steel as an absorber and quartz fibers as the sensitive material. The two halves of the HF are located 11.2\unit{m} from the interaction region, one on each end, and together they provide coverage in the range $3.0 < \abs{\eta} < 5.2$. They also serve as luminosity monitors. Each HF calorimeter consists of 432 readout towers, containing long and short quartz fibers running parallel to the beam. The long fibers run the entire depth of the HF calorimeter (165\cm, or approximately 10 interaction lengths), while the short fibers start at a depth of 22\cm from the front of the detector. By reading out the two sets of fibers separately, it is possible to distinguish showers generated by electrons and photons, which deposit a large fraction of their energy in the long-fiber calorimeter segment, from those generated by hadrons, which produce on average nearly equal signals in both calorimeter segments.

Events are selected online by means of a two-tiered trigger system~\cite{Khachatryan:2016bia}. 
The first level, composed of custom hardware processors, uses 
information from the calorimeters and muon detectors to select events 
at a rate of around 100\unit{kHz} within a fixed latency of less than 4\mus. 
The second level, known as the high-level trigger, consists of a 
farm of processors running a version of the full event reconstruction 
software optimized for fast processing, and reduces the event rate to 
around 1\unit{kHz} before data storage.

A more detailed description of the CMS detector, together with a definition of the coordinate system used and the relevant kinematic variables, is reported in Ref.~\cite{Chatrchyan:2008zzk}.

\section{Event selection}

The data sample used for the present analysis was collected with the CMS 
detector in 2013, when the LHC collided protons at $\sqrt{s} = 2.76\TeV$. The sample 
corresponds to an integrated luminosity of 5.4\pbinv.

Jets are clustered from particles reconstructed with the particle-flow (PF) 
algorithm~\cite{CMS-PRF-14-001}; this algorithm reconstructs and identifies each individual particle in an event, with an optimized combination of information from the various elements of the CMS detector. The vertex with the largest value of the summed charged-particle track 
\pt is the primary $\Pp\Pp$ interaction vertex (PV). The photon energy is obtained from the ECAL measurement. The electron energy 
is determined from a combination of the electron momentum at the PV as measured by the tracker, the energy of the 
corresponding ECAL cluster, and the energy sum of all bremsstrahlung 
photons spatially compatible with originating from the electron track. 
The muon energy is obtained from the curvature of the corresponding 
track. The energy of charged hadrons is determined from a combination of 
their momentum measured in the tracker and the matching ECAL and HCAL 
energy deposits, corrected for zero-suppression effects and for the 
response function of the calorimeters to hadronic showers. Finally, the 
energy of neutral hadrons is obtained from the corresponding corrected 
ECAL and HCAL energies. The silicon tracker measures charged particles within the range $\abs{\eta} < 2.5$. Therefore, jets with higher pseudorapidity are reconstructed using the calorimeter information only.

The clustering of jets is carried out with the anti-\kt algorithm~\cite{Cacciari:2008gp} as implemented in the \FASTJET package~\cite{Cacciari:2011ma} with a distance parameter 0.5. Jets in the analysis are required to have their charged tracks associated with the PV. Jet momentum is determined as the 
vector sum of all particle momenta in a jet, and is found from 
simulations to be within 5 to 10\% of the true momentum over the entire \pt 
spectrum and detector acceptance.

Additional $\Pp\Pp$ interactions (pileup) within the same or nearby bunch crossings can contribute additional tracks and calorimetric energy depositions, increasing the apparent jet momentum \cite{CMS:2020ebo}. To mitigate this effect, tracks identified as originating from pileup vertices are discarded, and an offset correction is applied to correct for remaining contributions. Jet energy corrections are derived 
from simulation, and are confirmed with in situ measurements of the 
energy balance in dijet and photon+jet, Z+jet, and multijet events. Additional selection 
criteria are applied to each event to remove spurious jet-like features 
originating from isolated noise patterns in certain HCAL regions.

The jet energy resolution (JER) typically amounts to 15\% at 10\GeV, 8\% at 100\GeV, and 4\% at 1\TeV~\cite{Khachatryan:2016kdb}.

Dijet events with moderate \Dy were selected online by 
means of a single jet trigger that required the presence of at least one 
jet with $\pt > 20\GeV$ and $\abs{\eta} < 5.0$. 
Dedicated single and double forward jet triggers were used 
to select dijet events with large \Dy. 
The single forward jet trigger required the presence of a jet with $\pt > 20\GeV$ 
in the forward or backward $\eta$ regions, \ie, $3.0 < \eta < 5.1$ or $-5.1 < \eta < -3.0$. Finally, the double forward jet trigger required at least one 
jet in the forward and at least one in the backward region with $\pt > 
20\GeV$. To keep the rate of the triggers within the allocated bandwidth, prescale factors were used, which reduced the effective integrated luminosities to 47.3\nbinv, 382\nbinv, and 4.6\pbinv for the single, single forward, and double forward jet triggers, respectively. The trigger efficiency for selecting events with at least one pair of jets with $\pt > 35\GeV$ is higher than 98\% for the single and single forward jet triggers, and is 100\% for the double jet trigger.

The combination of the samples obtained with the triggers just discussed 
provides the coverage of the kinematic region relevant for the present 
analysis. The overlap regions are removed, and the samples are reweighted and merged. The weights are calculated as the ratios of the effective
luminosities of the triggers. To exclude overlaps, appropriate offline
selection conditions are imposed. Specifically, the events containing
jets with $\pt > 35\GeV$ in the forward or backward
$\eta$ regions are removed from the single-jet trigger sample. Events
with at least one jet with $\pt > 35\GeV$ in the
forward or backward regions, but not in both, are kept in the sample
selected by the single forward jet trigger. In addition, the events selected
by the double forward jet trigger are required to have at least one jet
with $\pt > 35\GeV$, both in the forward and
backward regions. Control distributions are extracted from the single
jet trigger sample without the exclusion of the forward-backward events
described above. The event distributions from the combined sample coincide with the 
control distributions within statistical fluctuations.

Offline, events are required to have at least one reconstructed vertex. 
 The PV must contain at least four tracks associated with it, should be within 2\cm of 
the beam axis in the transverse plane, and its $z$ coordinate should be 
within the luminous collision region, $\abs{z_\mathrm{PV}} < 24\unit{cm}$. The efficiency for selecting events with at least one vertex is 98.9\% for the single jet triggers and 99.5\% for the double jet trigger. Background originating from the beam scraping some aperture limitation in the 
beamline is suppressed by requiring that events with more than ten 
tracks have a fraction of high-purity tracks of at least 25\%~\cite{Chatrchyan:2014fea}.

The events of interest for the present analysis are those that contain 
at least two jets with $\pt > 35\GeV$ and $\abs{y} < 4.7$; jets with $\pt > \ptveto = 20\GeV$ are considered for the ``exclusive" with veto sample.

\section{Simulated samples}

The simulations are used to compare the results with
theoretical models and to correct for detector effects. The following MC generators are used: \PYTHIAeight (8.183)~\cite{Sjostrand:2007gs}, tune 4C~\cite{Corke:2010yf} and 
\HERWIGpp (2.7.1)~\cite{Bahr:2008pv}, tune EE3C~\cite{Gieseke:2011na}. These both implement leading-order (LO) matrix 
element predictions improved with LL DGLAP-based 
parton showering. \POWHEG (2.0)~\cite{Alioli:2010xa} generates events according to 
NLO matrix element predictions; parton showering 
is carried out with \PYTHIAeight (8.230), \HERWIGseven (7.1.2)~\cite{Bellm:2015jjp}, and 
\HERWIGpp (2.7.1). LL BFKL-based predictions are provided 
by \HEJ (1.4.0)~\cite{Andersen:2011hs} at parton level; the hadronization is carried out 
with the \ARIADNE (4.12J01) generator~\cite{Lonnblad:1992tz}.

\section{Corrections for detector effects}\label{sect:corrections_det_effects}

Trigger efficiencies are measured as functions of $\eta$ and \pt using a 
minimum bias sample. The single jet trigger is 92\% efficient in 
selecting events with \pt of the leading jet greater than 35\GeV. The single jet 
trigger is also used for the determination of the single forward jet 
trigger efficiency, which is at least as efficient as the single jet trigger. The double forward jet trigger is 100\% efficient in 
selecting dijet events with the two leading jets on opposite sides. Therefore, only the single jet trigger inefficiency needs to be corrected. Dijet events have higher probability of being selected by the single jet trigger because each jet in the event can satisfy the 
trigger conditions. The total impact of the trigger inefficiency 
correction does not exceed 0.16\% for the cross section ratios and 1.5\% 
for the cross sections.

The finite resolution of the reconstructed values of the jet \pt and $y$ leads to migration of the events across the \pt threshold and between $y$ bins, resulting in distortions of the 
measured distributions. To study the impact of these effects and correct for them, MC 
simulations are employed. 

Events at stable-particle level generated with \PYTHIAeight
and \HERWIGpp are passed through the standard \GEANTfour~\cite{Agostinelli:2002hh} 
simulation of the CMS detector and reconstructed in the same way as the 
collision data. The PU contribution is included by adding simulated $\Pp\Pp$ interactions with a rate equal to that observed in data. A study in Ref.~\cite{Chatrchyan:2011ds} has shown that the JER for 
PF jets in the MC simulation is 7.9--40.0\% better than that 
in the data. To correct for this difference a further energy smearing 
is applied to the four-momentum of the simulated jets at the detector 
level.

To study the migrations of events in \pt and $y$, jets at stable-particle level are matched to jets at detector level. Matching requires a jet at stable-particle level to be within a 0.4 radius in $(\eta, \phi)$ space around the axis of the jet at detector level. The main distortion is caused by the migration across the \pt threshold. The migration from below to above the threshold causes 
background events to enter the sample. The background fraction increases 
 from 40\% at $\Dy < 1$ to 85\% at $\Dy > 7$. The migration from above to below the threshold leads to the loss 
of up to 40\% of the signal events. The migration between $y$ bins affects less than 10\% of the
events. Since migration effects are correlated across dijet samples, their impact on the ratio of dijet cross sections is reduced.

The migration effects are corrected by unfolding the data with the 
Tikhonov regularisation scheme, as implemented in the TUnfold~\cite{Schmitt:2012kp} 
package. 

{\tolerance=1000 As a consistency check, we verified that unfolding the detector level
distributions obtained with \PYTHIAeight (\HERWIGpp) by using response 
matrices also determined with \PYTHIAeight (\HERWIGpp) gives distributions 
consistent with the generated ones. Conversely, if the \PYTHIAeight distributions are unfolded with \HERWIGpp, residual differences of up to 20\% are found for the cross section measurement and 4.4\% for the ratios; the residuals are approximately the same if the \HERWIGpp distributions are unfolded with \PYTHIAeight.
These residual differences do not change when both \HERWIGpp and \PYTHIAeight simulations are reweighted to match the data at the detector level. The bias produced by the change in shape of the reweighted simulation used for the unfolding is fully covered by the residual difference. The residual differences reflect the fact that the detector response to jets is simulated with different models. Therefore, the difference in the results obtained by unfolding the data with different models is an estimate of the model dependence of the unfolding procedure.\par}

The average unfolding correction obtained with \PYTHIAeight and \HERWIGpp is 
used for the results. The unfolding corrections are in the range 0.92--0.20 
for the differential cross section measurement, 0.97--1.01 
for the ratios without veto, and 0.95--0.98 
for 
the ratios with veto. The ranges correspond to the variation of the unfolding correction with increasing \Dy.

\section{Systematic uncertainties}

The following sources of systematic uncertainty are considered.
\begin{enumerate}
\item \textit{Jet energy scale (JES) uncertainty.} Its effect is estimated by scaling the four-momentum of the jets by the \pt and $\eta$ dependent uncertainty of the jet energy correction factors, which are determined by using the multistep approach described in Ref.~\cite{Chatrchyan:2011ds}. The magnitude of the uncertainty does not exceed 4.6 (3.2)\% for jets with $\pt = 20$ (35)\GeV. The difference between the nominal results and those obtained with the scaling is an estimate of the effect of the JES uncertainty in the results.

\item \textit{Uncertainty in the JER correction factors.} The systematic uncertainty of the MC modeling of the JER depends on $\eta$ of the jets~\cite{Chatrchyan:2011ds}, and varies between 2.6 and 19.0\%. The contribution of this uncertainty is evaluated by smearing the 
four-momentum of the simulated jets at the detector level according to 
the JER correction uncertainty. The difference between the nominal results and those with the additional smearing is an estimate of the systematic uncertainty.

\item \textit{Model dependent (MD) uncertainty of the unfolding.} The difference between the results obtained correcting the data with \PYTHIAeight and \HERWIGpp is an estimate of the model dependence of the unfolding. This uncertainty varies between 1.5 and 22.0\% for the cross sections, whereas it is between 0.5 and 4.5\% for the ratios.

\item {\tolerance=1050 \textit{Uncertainty related to the parton distribution function (PDF) set.} The \PYTHIAeight (tune 4C) and \HERWIGpp (tune EE3C) samples used for the unfolding employ the LO CTEQ6L1 PDF set~\cite{Pumplin:2002vw}. The PDF uncertainty is propagated as recommended by the PDF4LHC group~\cite{Butterworth:2015oua}. The following PDF sets are used for the propagation: NNPDF31\_lo\_as\_0130~\cite{Ball:2014uwa}, CT14lo~\cite{Dulat:2015mca}, and MMHT2014lo68cl~\cite{Harland-Lang:2014zoa}. The PDF reweighting scheme is described in Ref.~\cite{Buckley:2014ana}. The reweighted MC distributions and response matrices obtained are used for the unfolding. The difference between the nominal results and the envelope of those obtained with the reweighted MC is an estimate of the uncertainty due to the PDFs.\par}

\item \textit{Renormalization and factorization scales uncertainty.} The uncertainty related to the renormalization and factorization scales, \mur and \muf, of the QCD renormalization group equation is estimated by 
modifying \mur and \muf independently by factors of 2 and 0.5 in \PYTHIAeight. The cases where one scale is simultaneously increased and the other is decreased by a factor 2 are not considered. The difference between the nominal results and the envelope of those obtained with the modified scales is a contribution to the systematic uncertainty.

\item \textit{Uncertainty because of the limited MC statistics (MCS).} The propagation of the statistical uncertainty in the simulation is carried out with the statistical bootstrap method. 

\item \textit{The uncertainty in the integrated luminosity} is 3.7\%~\cite{CMS-PAS-LUM-13-002}. It is treated as a normalization uncertainty for the cross section measurements. The luminosity uncertainty cancels in the cross section ratios.

\item \textit{Uncertainty in the trigger efficiency corrections (TEC).} Trigger efficiencies are investigated as functions of $\eta$ and \pt. The impact of the trigger inefficiency does not exceed 0.16\% for the cross section ratios and 1.5\% for the cross sections. The uncertainty in the correction is estimated as 100\% of the correction itself.

\item {\tolerance=800 \textit{Uncertainty related to PU.} The sensitivity of the results to PU is investigated by subdividing
the data sample into low- and high-PU subsamples depending on the
instantaneous luminosity; no dependence of the results on PU is observed. As mentioned earlier, the MC samples are weighted to match the amount of PU observed in the data and the weighted samples are used for the unfolding. The difference between the results obtained with the weighted and unweighted samples is smaller than 0.05\% for the ratios without veto, 0.96\% for the ratios with the veto, and 3.2\% for the cross section measurement. The systematic uncertainty related to the weighting procedure is estimated to be equal to these differences. \par}

\end{enumerate}

\begin{figure}[p!]
  \centering
    \includegraphics[width=0.47\textwidth]{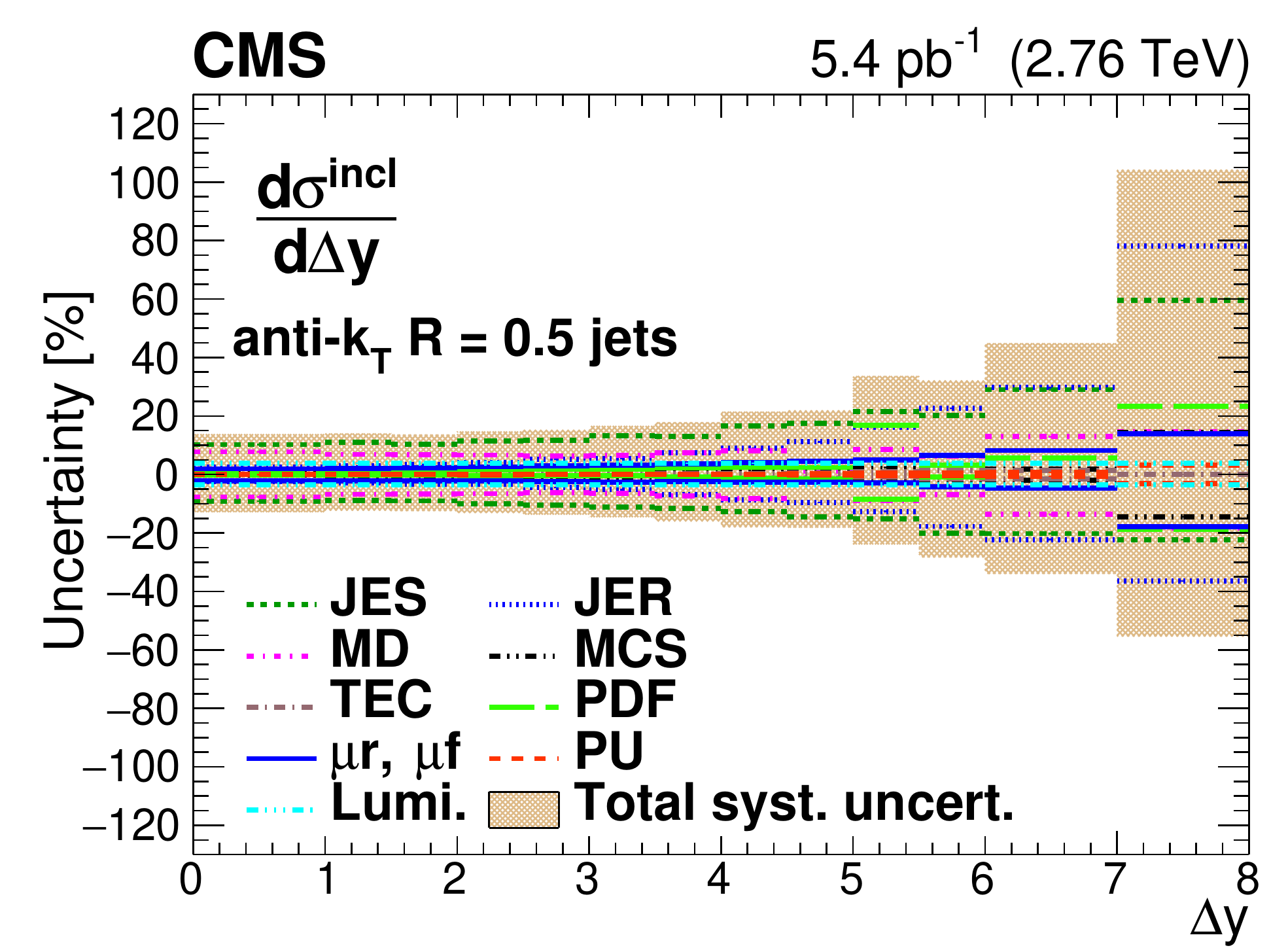}
    \includegraphics[width=0.47\textwidth]{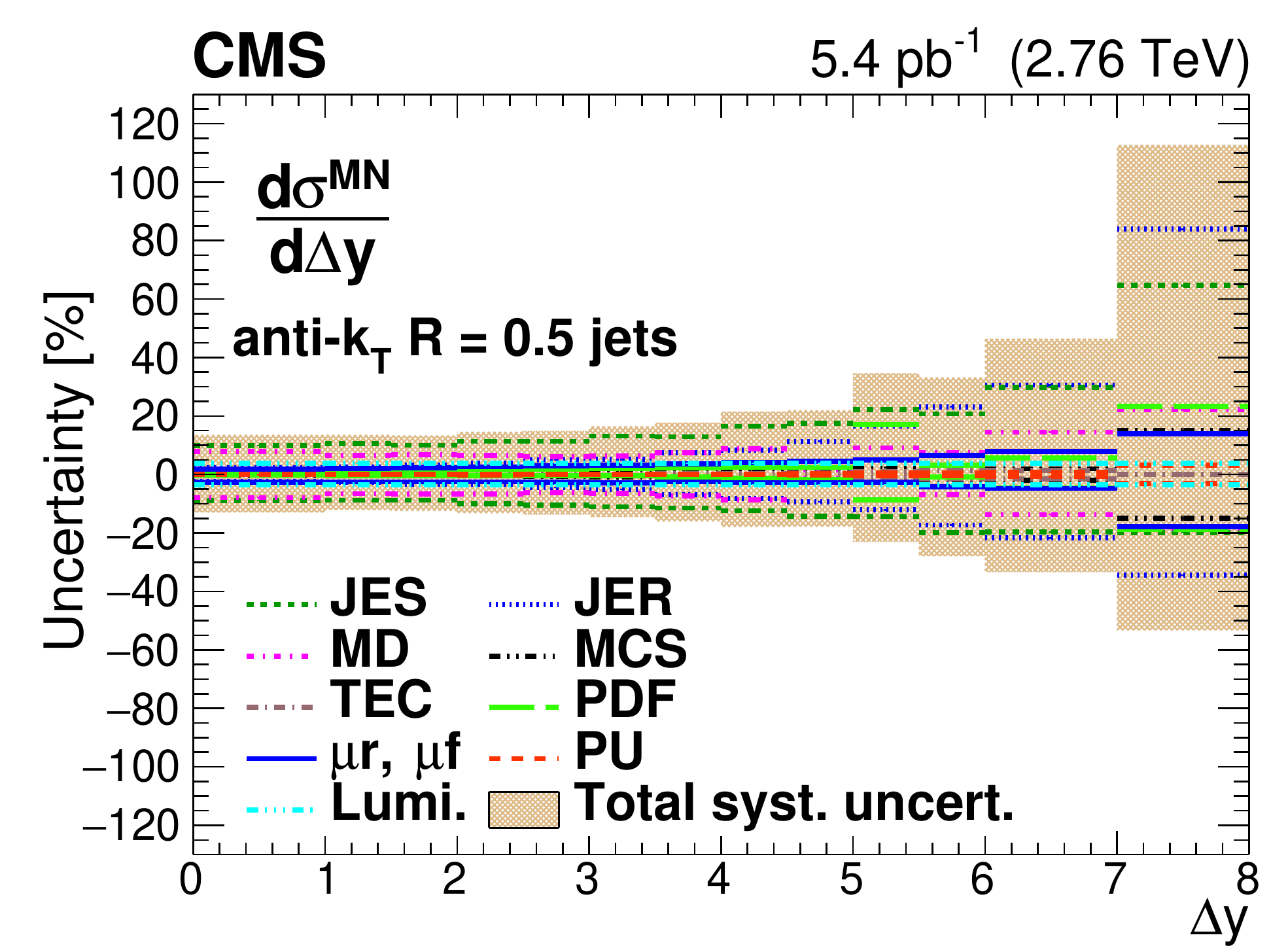}
    \includegraphics[width=0.47\textwidth]{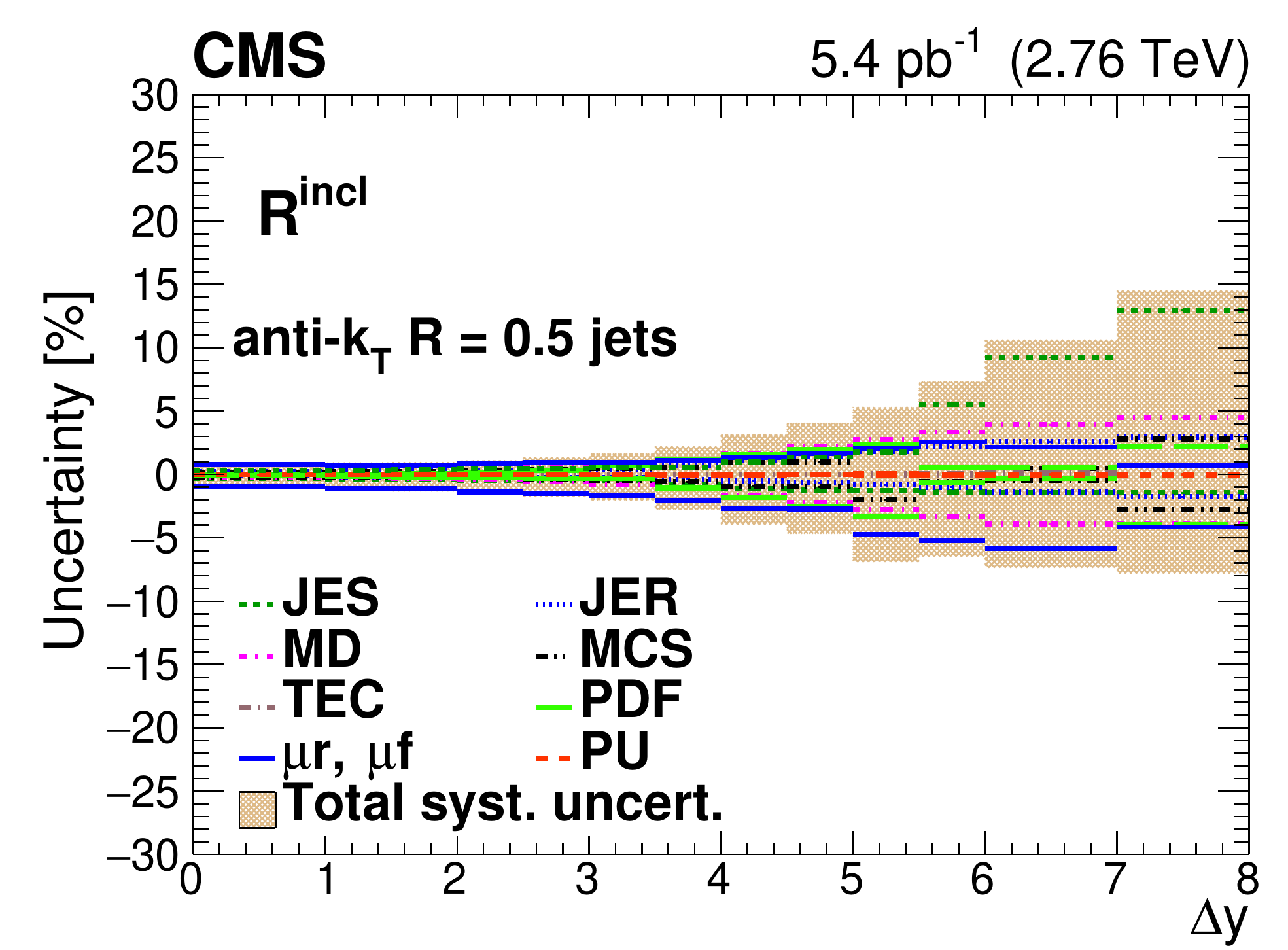}
    \includegraphics[width=0.47\textwidth]{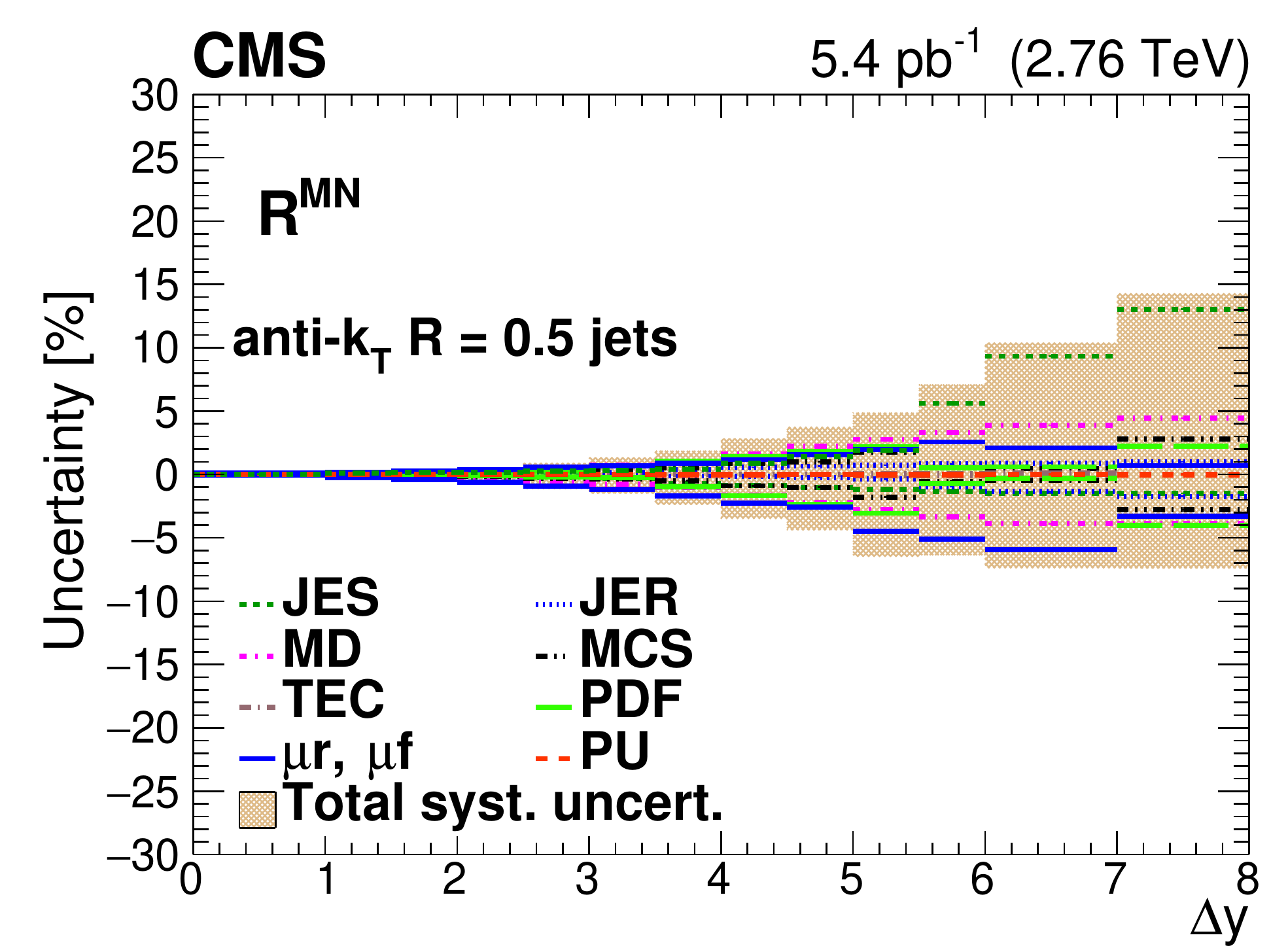}
    \includegraphics[width=0.47\textwidth]{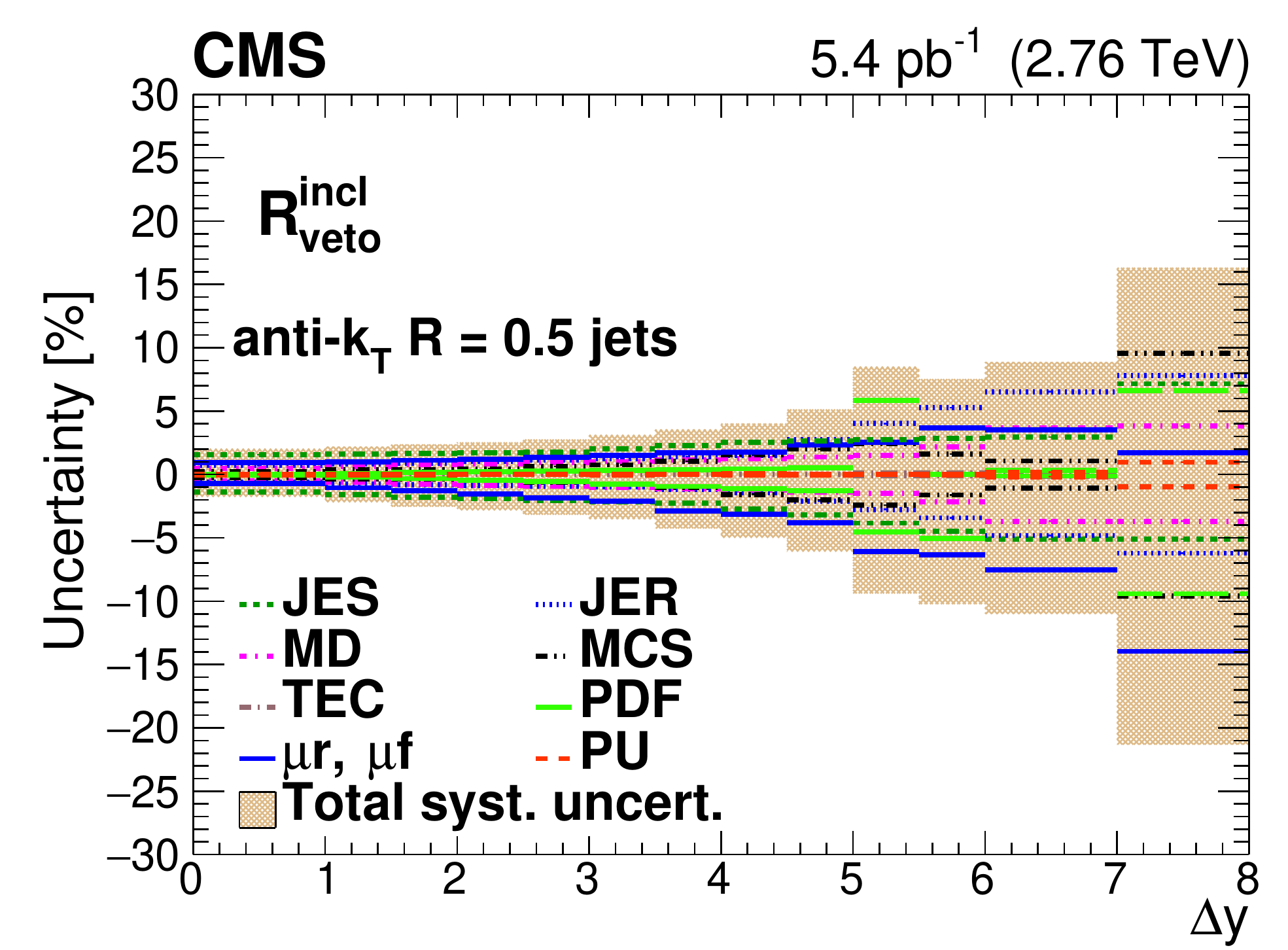}
    \includegraphics[width=0.47\textwidth]{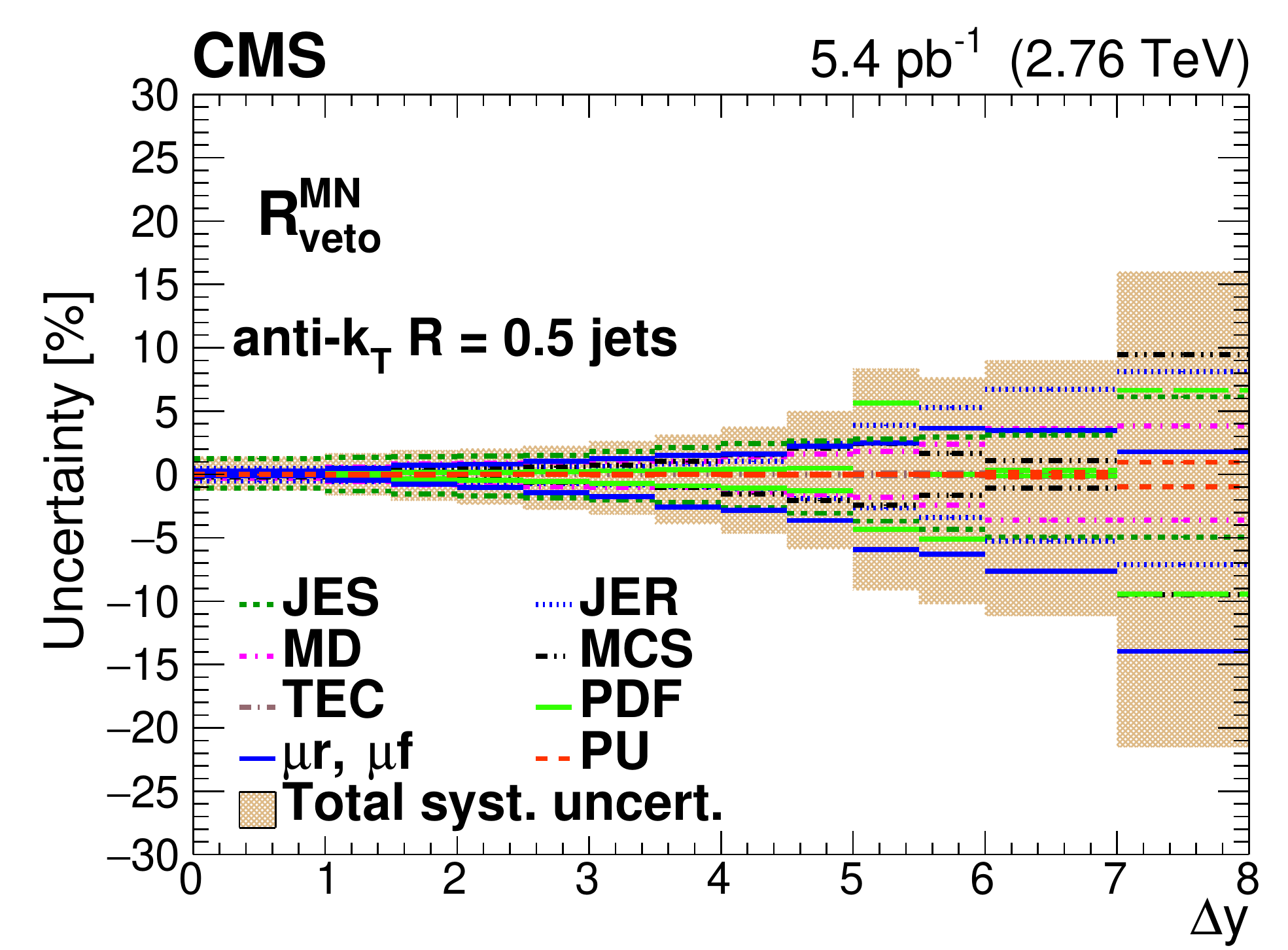}
    \caption{Summary of the systematic uncertainties on the cross sections $\ddinline{\sgincl}{\Dy}$ (upper left) and $\ddinline{\sgmn}{\Dy}$ (upper right), as well as the ratios $\Rincl$ (middle left), $\RMN$ (middle right), $\Rinclveto$ (lower left), and $\RMNveto$ (lower right). The various contributions are indicated by the lines and the total
uncertainty is shown with a band.}
    \label{fig:unc_ak5}

\end{figure}

The contributions just discussed are shown in Fig.~\ref{fig:unc_ak5}. The correlations between the uncertainty sources in the numerator and denominator of the ratios in Eq.~(\ref{eq:k-factors}) lead to significant improvements with respect to the individual cross section measurements. The dominant contributions to the systematic uncertainty in the cross section 
measurements are given by the JES and JER uncertainties. However, systematic uncertainties from different sources are approximately of the same order for the ratios, except the uncertainty in the trigger inefficiency and PU, which is much smaller than the other contributions. The total systematic uncertainty is the quadratic 
sum of all its components.

\section{Results}
Cross sections for inclusive and MN dijet production 
are presented, along with various cross section ratios for jets with $\pt > 35\GeV$ and $\abs{y} < 4.7$. The results are compared to DGLAP-based and LL BFKL-based MC models.

\subsection{Differential cross sections \texorpdfstring{$\ddinline{\sgincl}{\Dy}$}{sgincDy} and \texorpdfstring{$\ddinline{\sgmn}{\Dy}$}{sgmnDy}}

\begin{figure}[hbt!]
  \centering
    \includegraphics[width=0.47\textwidth]{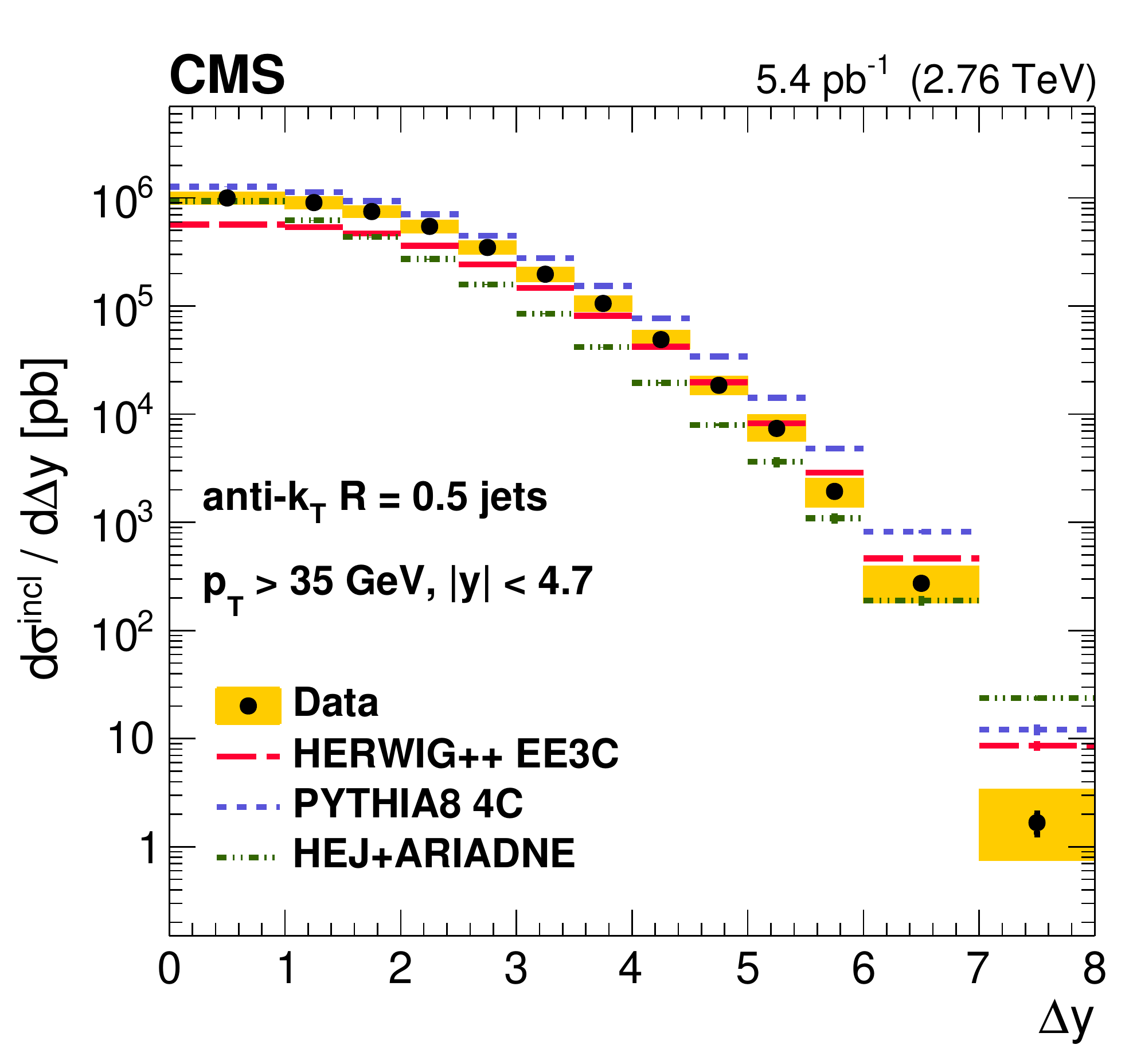}
    \includegraphics[width=0.47\textwidth]{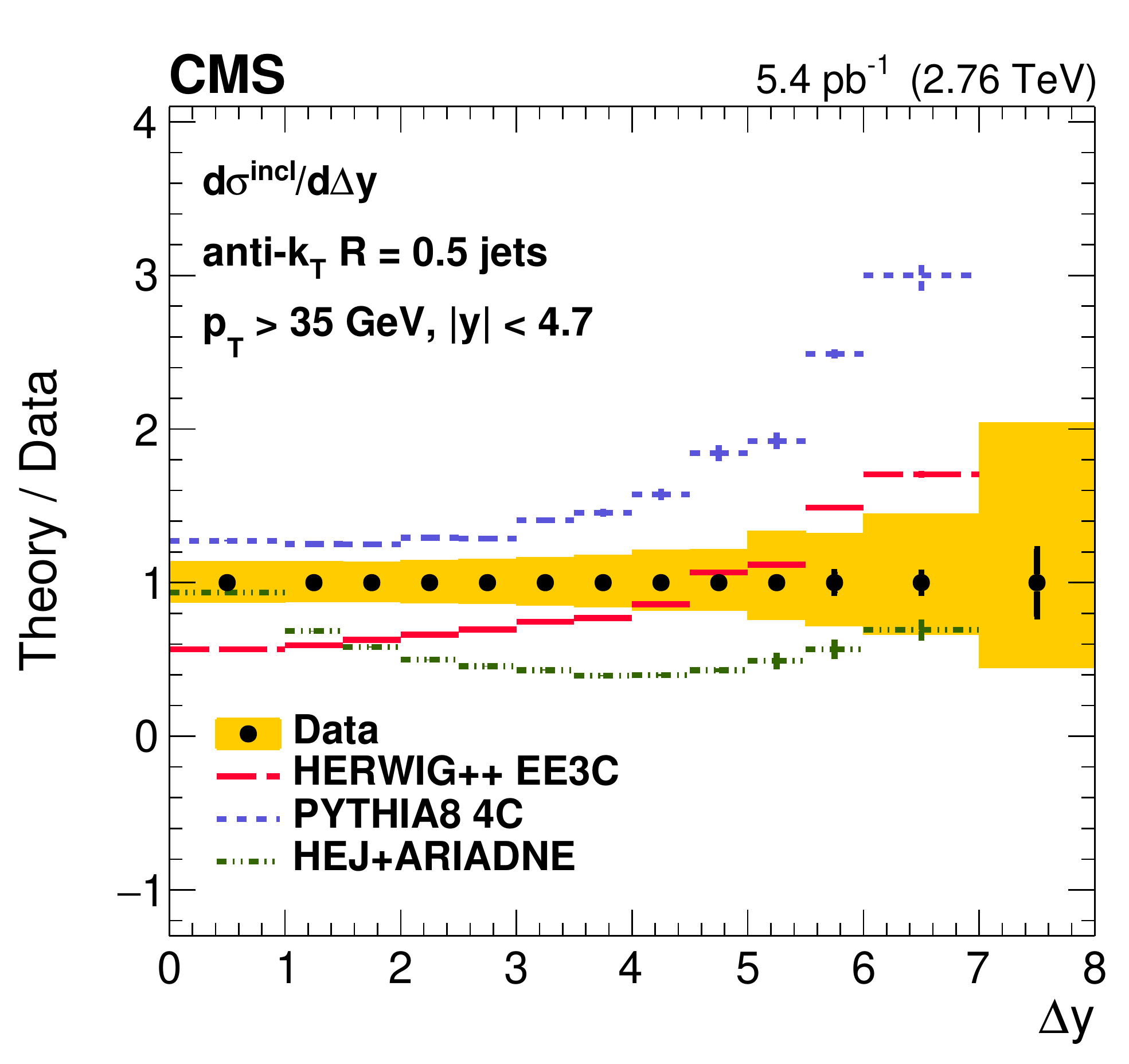}
    \includegraphics[width=0.47\textwidth]{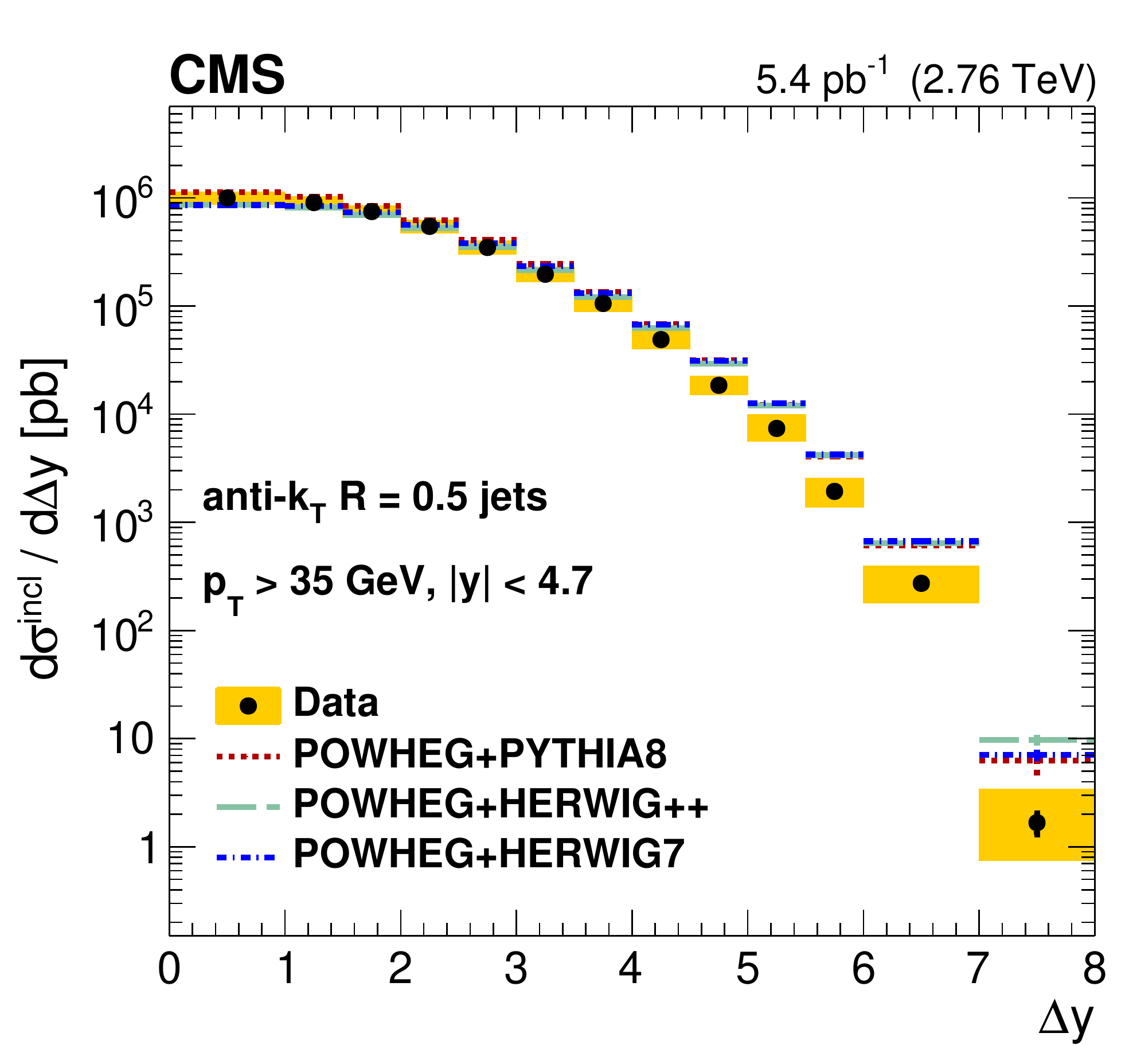}
    \includegraphics[width=0.47\textwidth]{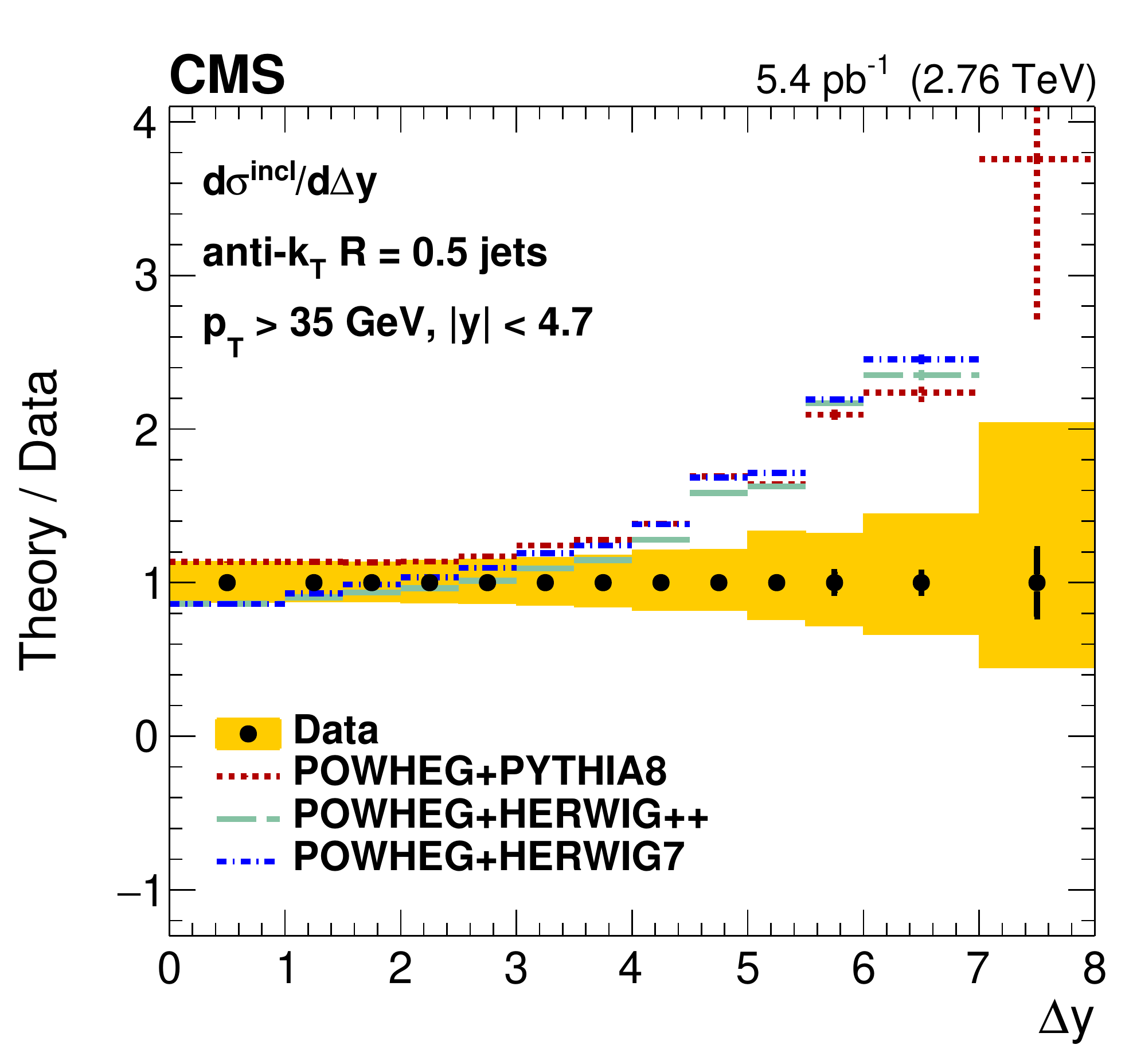}
    \caption{Differential cross section $\ddinline{\sgincl}{\Dy}$ for inclusive dijet production. The upper and lower rows present the comparison with different MC models. The plots on the left present the cross sections and those on the right the ratio of theory to data. The systematic uncertainties are indicated by the shaded bands and the statistical uncertainties by the vertical bars.}
    \label{fig:x_section_incl_ak5}
\end{figure}

The differential cross section $\ddinline{\sgincl}{\Dy}$ for inclusive dijet 
production is shown in Fig.~\ref{fig:x_section_incl_ak5}, together with the predictions of various 
models. The LO+LL DGLAP-based \PYTHIAeight simulation overestimates the data, whereas \HERWIGpp underestimates them for $\Dy < 4$ and overestimates them for $\Dy > 6$. The predictions of the LL BFKL-based calculation of \HEJARIADNE 
systematically underestimate the data for $\Dy < 7$ and overestimate them in the largest \Dy bin. The inclusion of the NLO 
corrections provided by \POWHEG improves the DGLAP-based 
predictions in the $\Dy < 4$ region. However, \POWHEGPYTHIAeight still overestimate the results. The predictions of \POWHEGHERWIG are consistent with the data for $\Dy < 4$, but overestimate them for 
large values of \Dy. The measured cross section decreases with 
\Dy faster than the DGLAP-based MC predictions.

\begin{figure}[hbt!]
  \centering
    \includegraphics[width=0.47\textwidth]{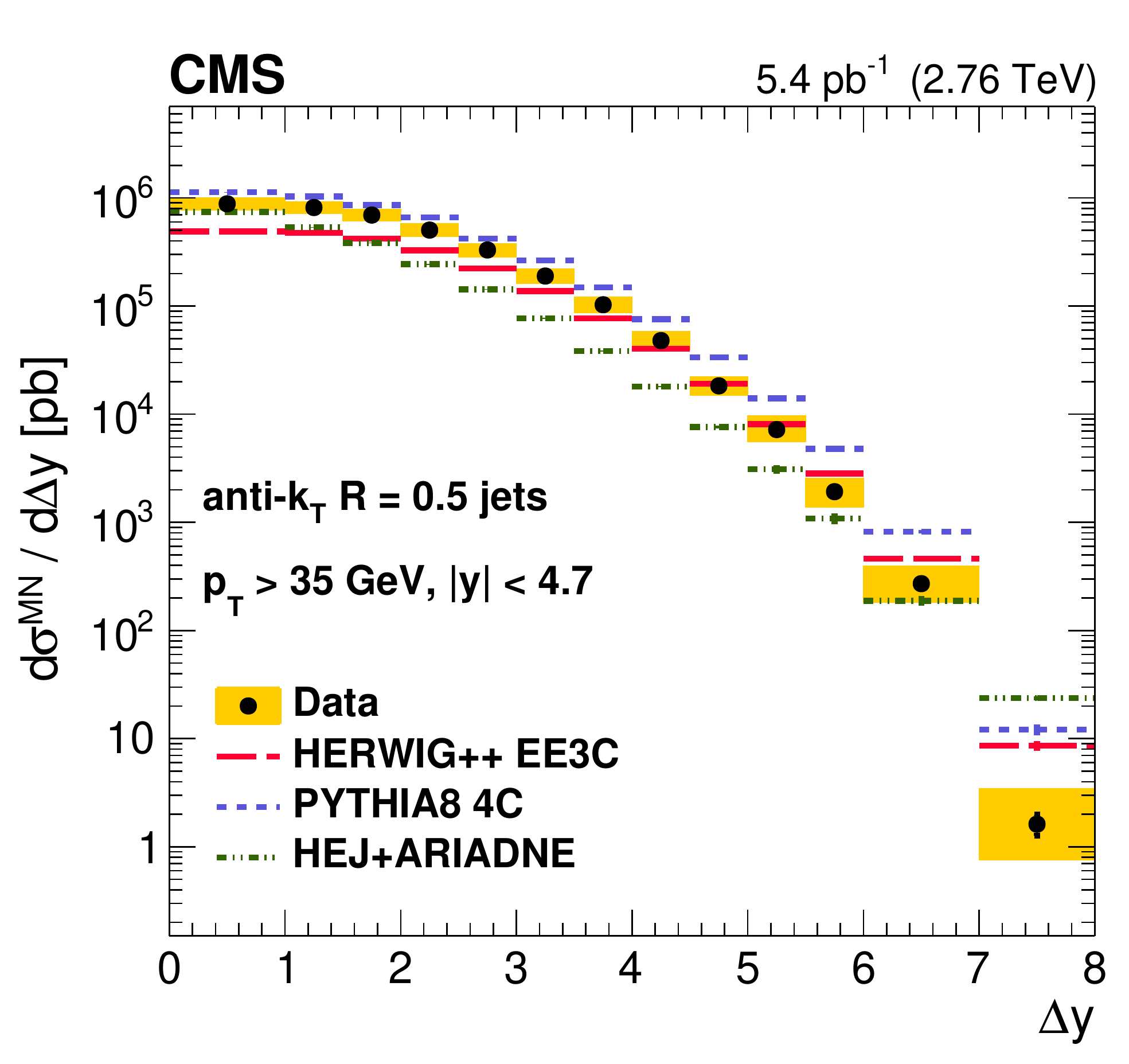}
    \includegraphics[width=0.47\textwidth]{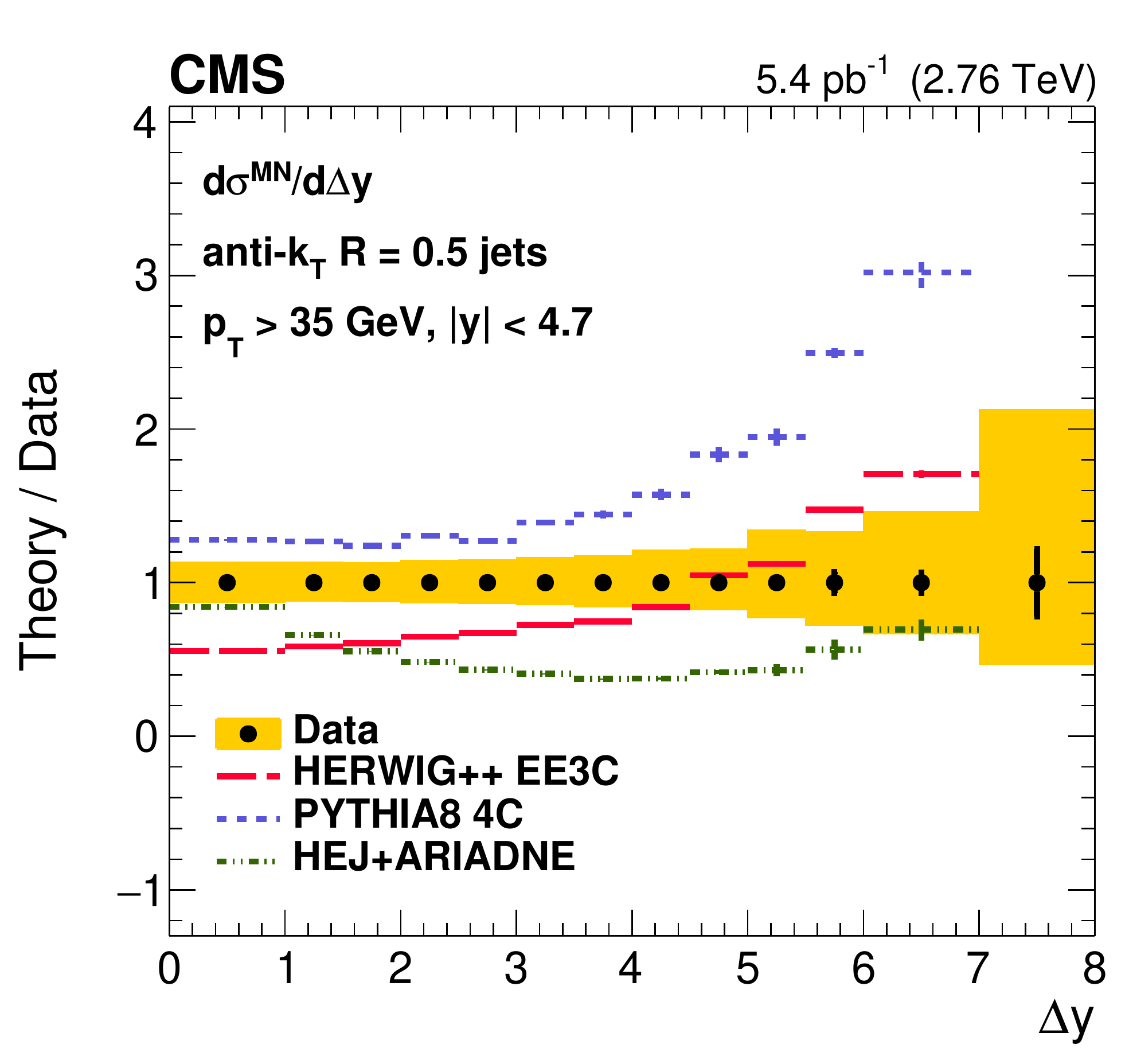}
    \includegraphics[width=0.47\textwidth]{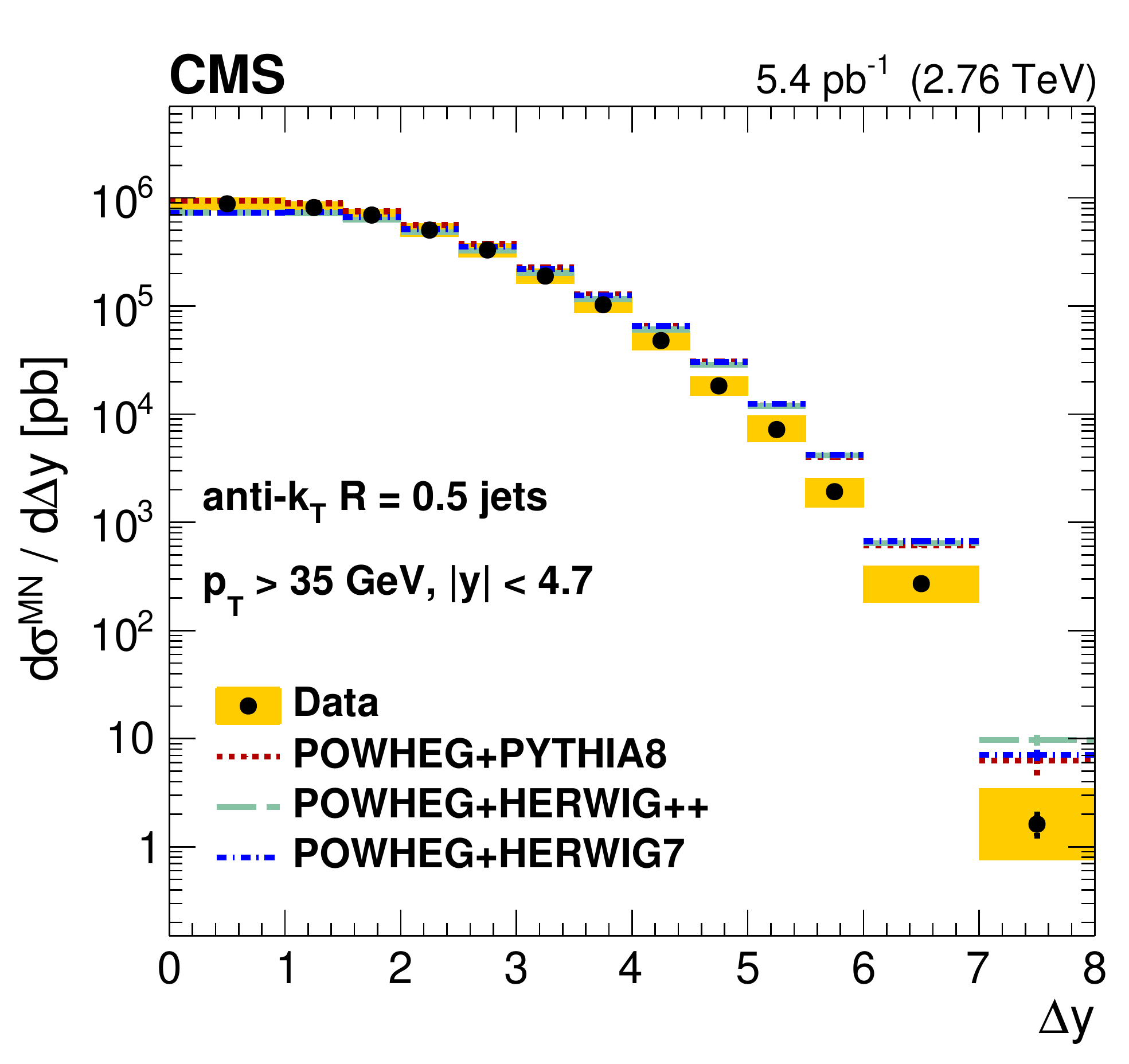}
    \includegraphics[width=0.47\textwidth]{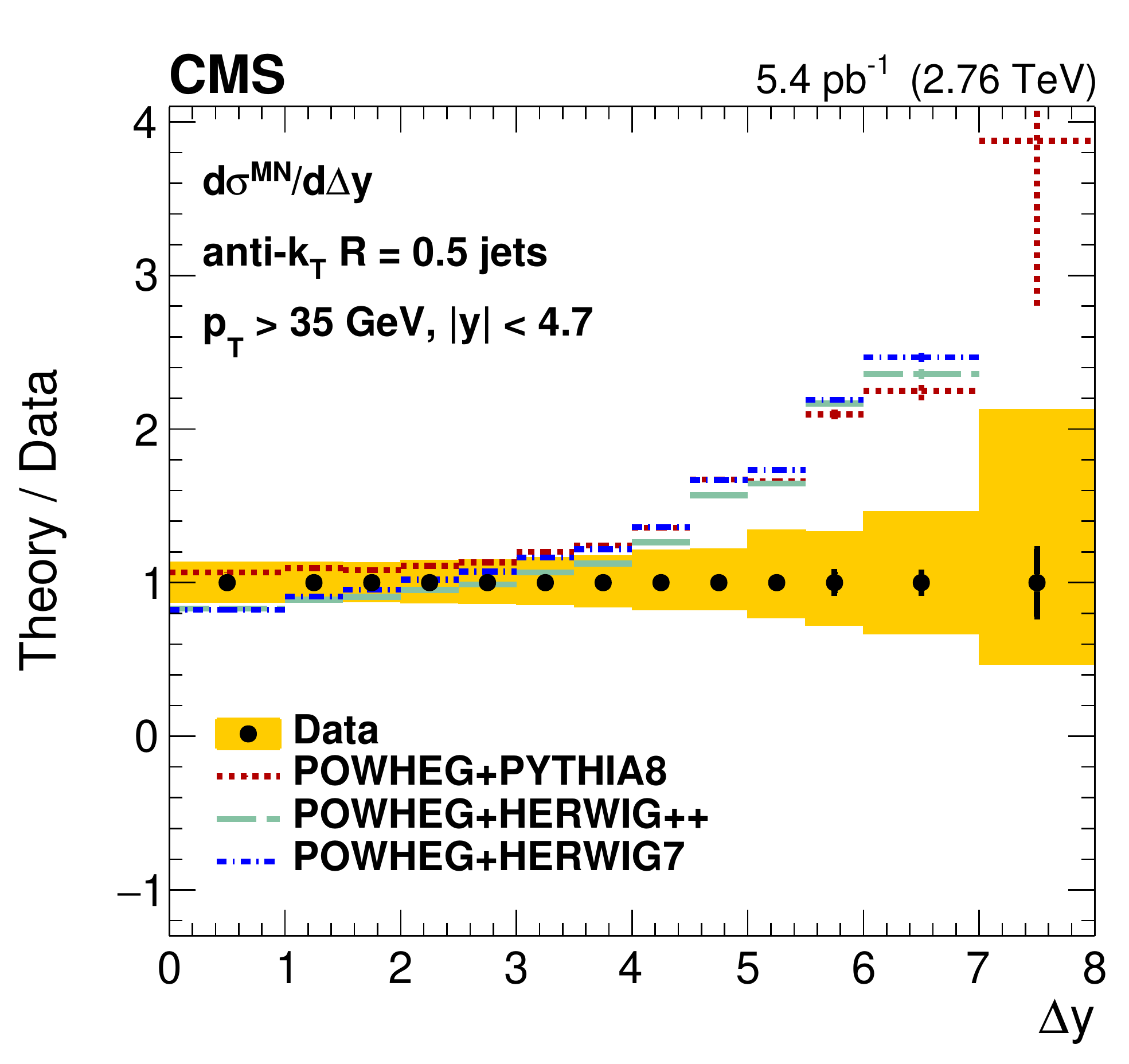}
    \caption{Differential cross section $\ddinline{\sgmn}{\Dy}$ for MN dijet production. The upper and lower rows present the comparison with different MC models. The plots on the left present the cross sections and those on the right the ratio of theory to data. The systematic uncertainties are indicated by the shaded bands and the statistical uncertainties by the vertical bars.}
    \label{fig:x_section_MN_ak5}
\end{figure}

The differential cross section for the production of MN dijets, $\ddinline{\sgmn}{\Dy}$, is presented in Fig.~\ref{fig:x_section_MN_ak5}. The data are compared 
with the same models used in Fig.~\ref{fig:x_section_incl_ak5}. The degree of agreement between data and the predictions is similar to that for the inclusive cross section discussed above.

\subsection{Ratio \texorpdfstring{\Rincl}{Rincl}}

\begin{figure}[hbt!]
  \centering
    \includegraphics[width=0.47\textwidth]{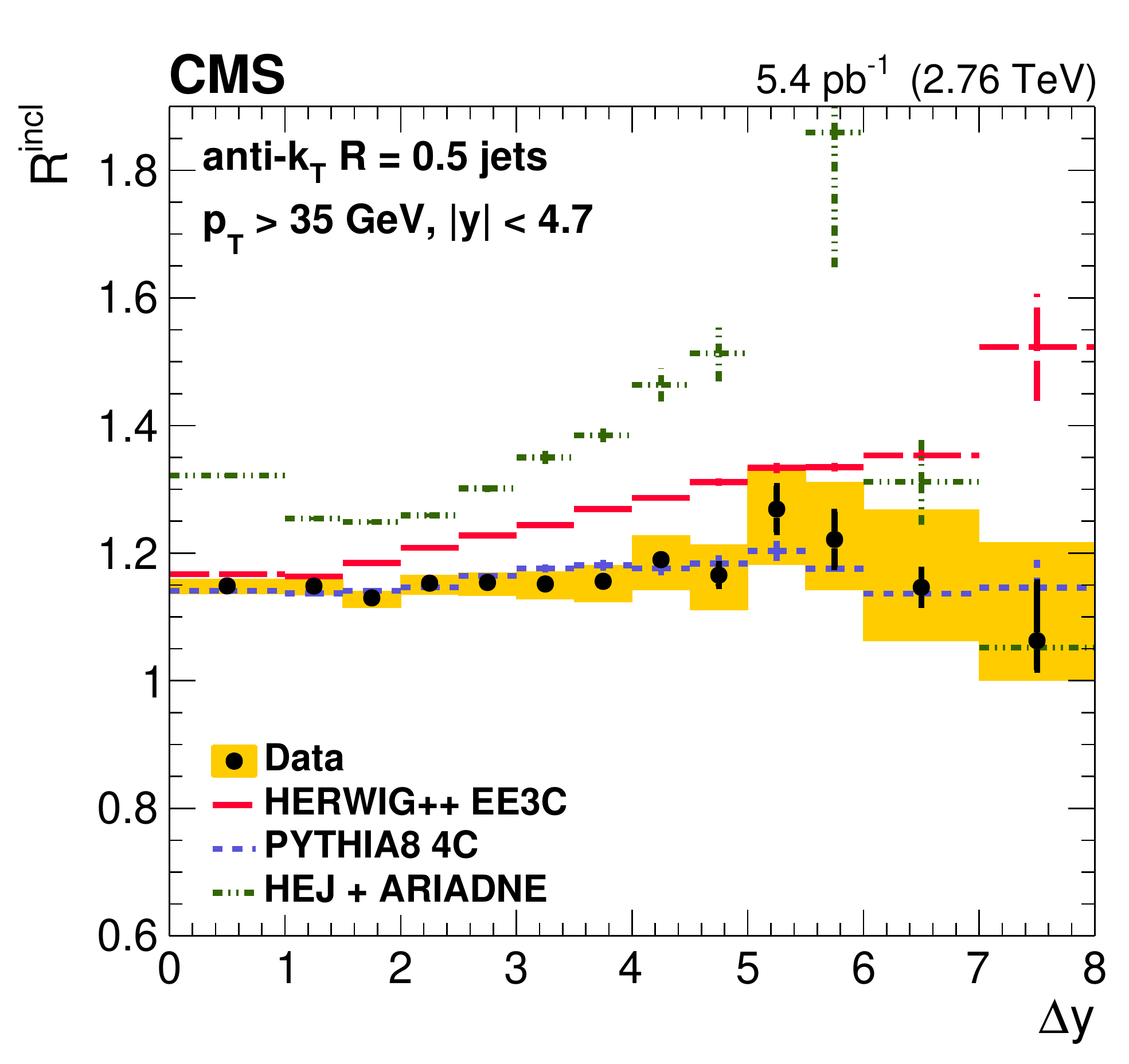}
    \includegraphics[width=0.47\textwidth]{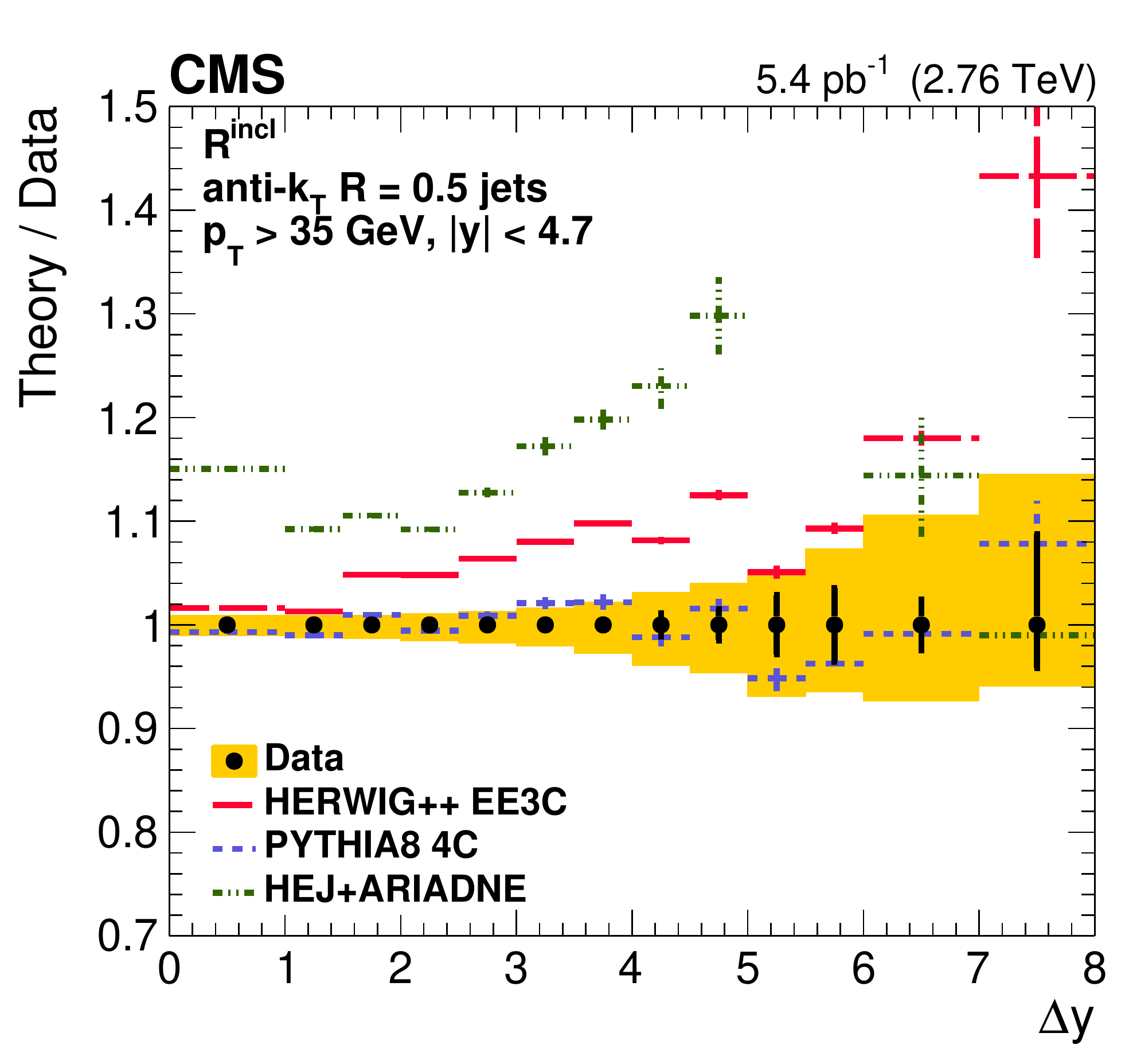}
    \includegraphics[width=0.47\textwidth]{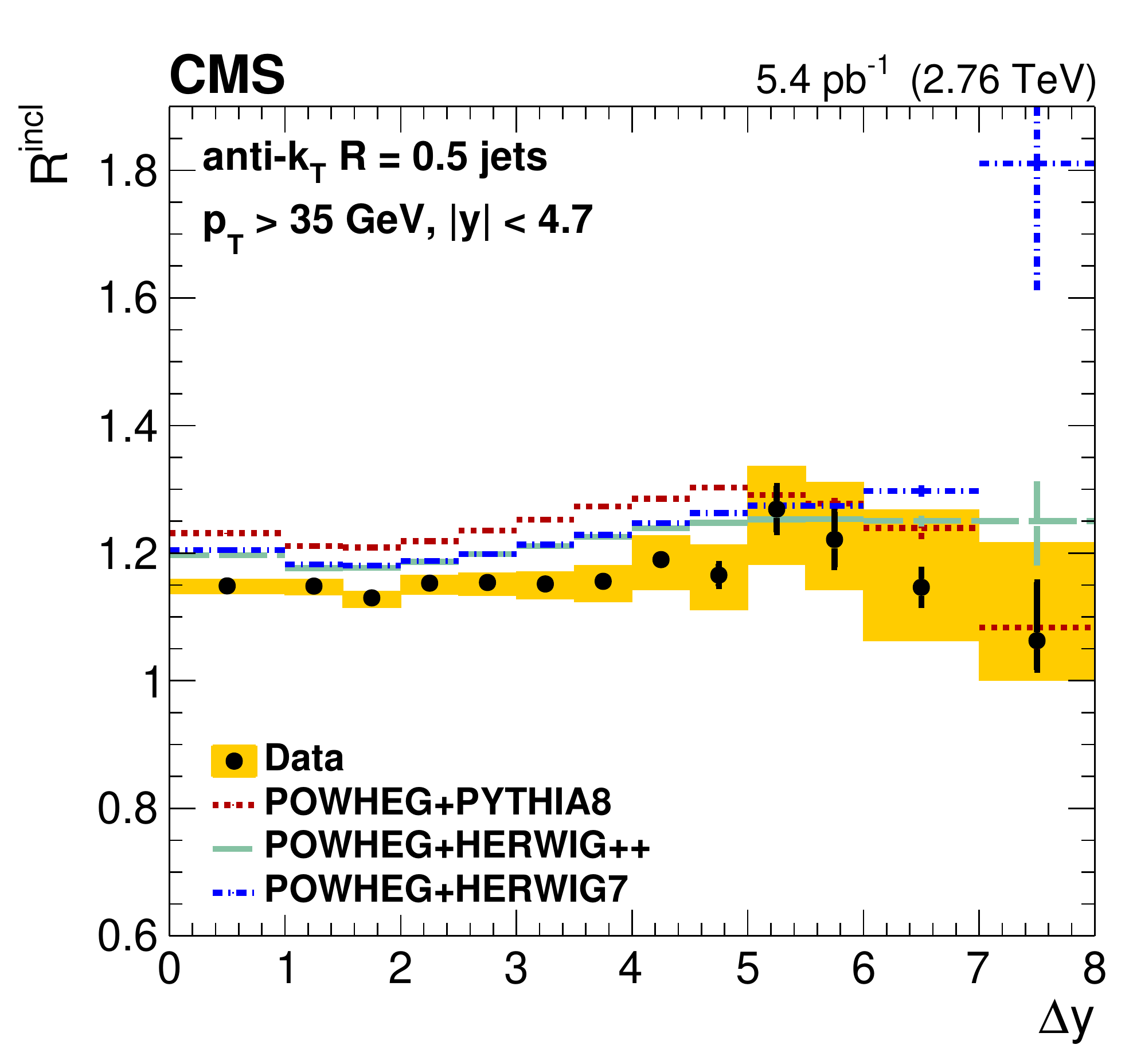}
    \includegraphics[width=0.47\textwidth]{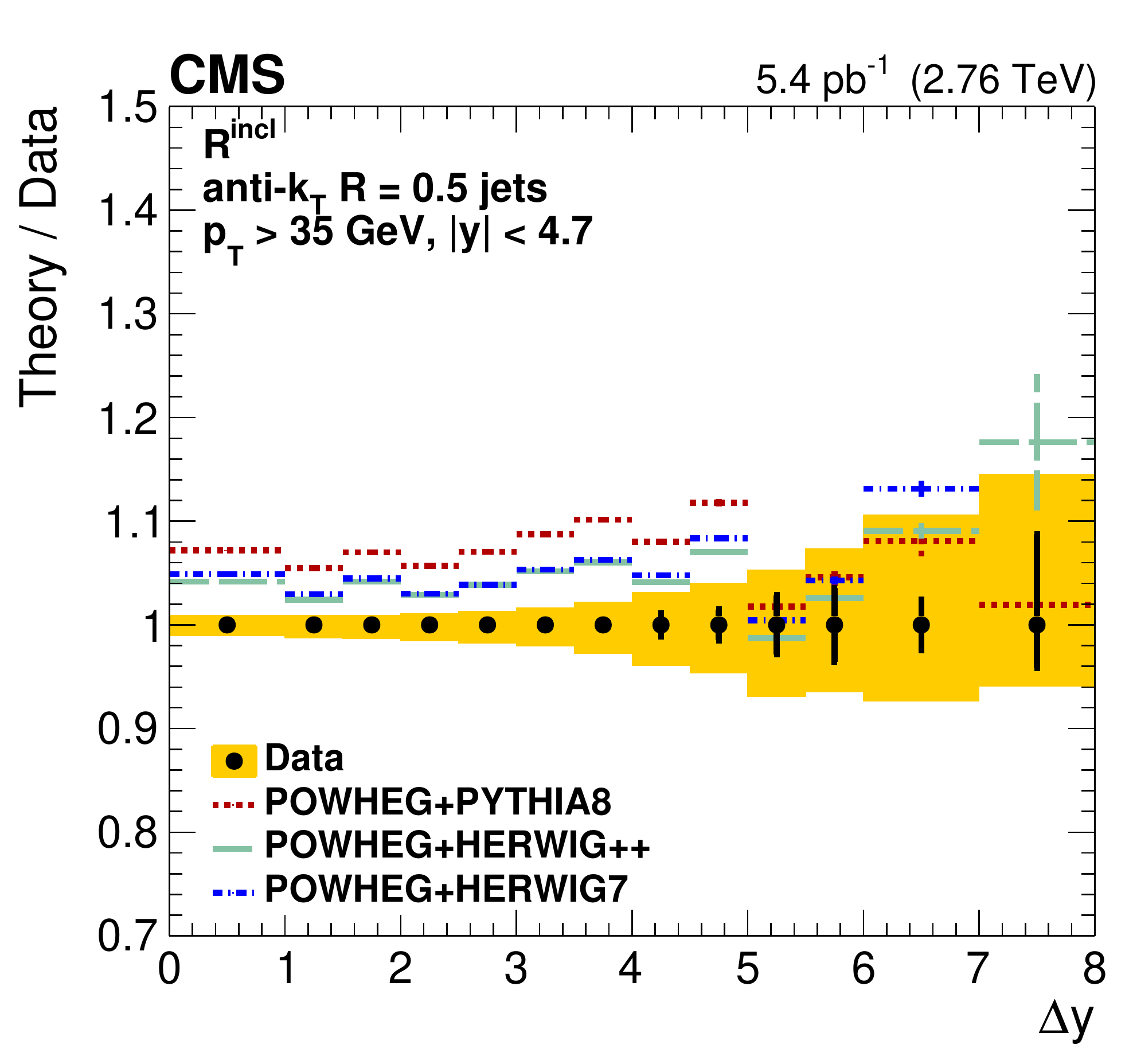}
    \caption{Ratio \Rincl of the cross sections for inclusive to ``exclusive" dijet production. The upper and lower rows present the comparison with different MC models. The plots on the left present the ratio \Rincl and those on the right the ratio of theory to data. The systematic uncertainties are indicated by the shaded bands and the statistical uncertainties by the vertical bars.}
    \label{fig:k_factor_incl_ak5}
\end{figure}

The ratio of the inclusive to the ``exclusive" dijet production cross 
sections, $\Rincl$, is presented in Fig.~\ref{fig:k_factor_incl_ak5}. The 
prediction of \PYTHIAeight is in agreement with the data, whereas \HERWIGpp 
and \HEJARIADNE overestimate the ratio. The addition of the NLO 
corrections via \POWHEG to the LL DGLAP-based prediction of \HERWIG 
degrades the quality of the description for $\Dy < 2$ but improves it 
for large rapidity intervals. The inclusion of the NLO corrections to 
\PYTHIAeight with \POWHEG worsens the agreement significantly.

\subsection{Ratio \texorpdfstring{\Rinclveto}{Rinclveto}}
\begin{figure}[hbt!]
  \centering
    \includegraphics[width=0.47\textwidth]{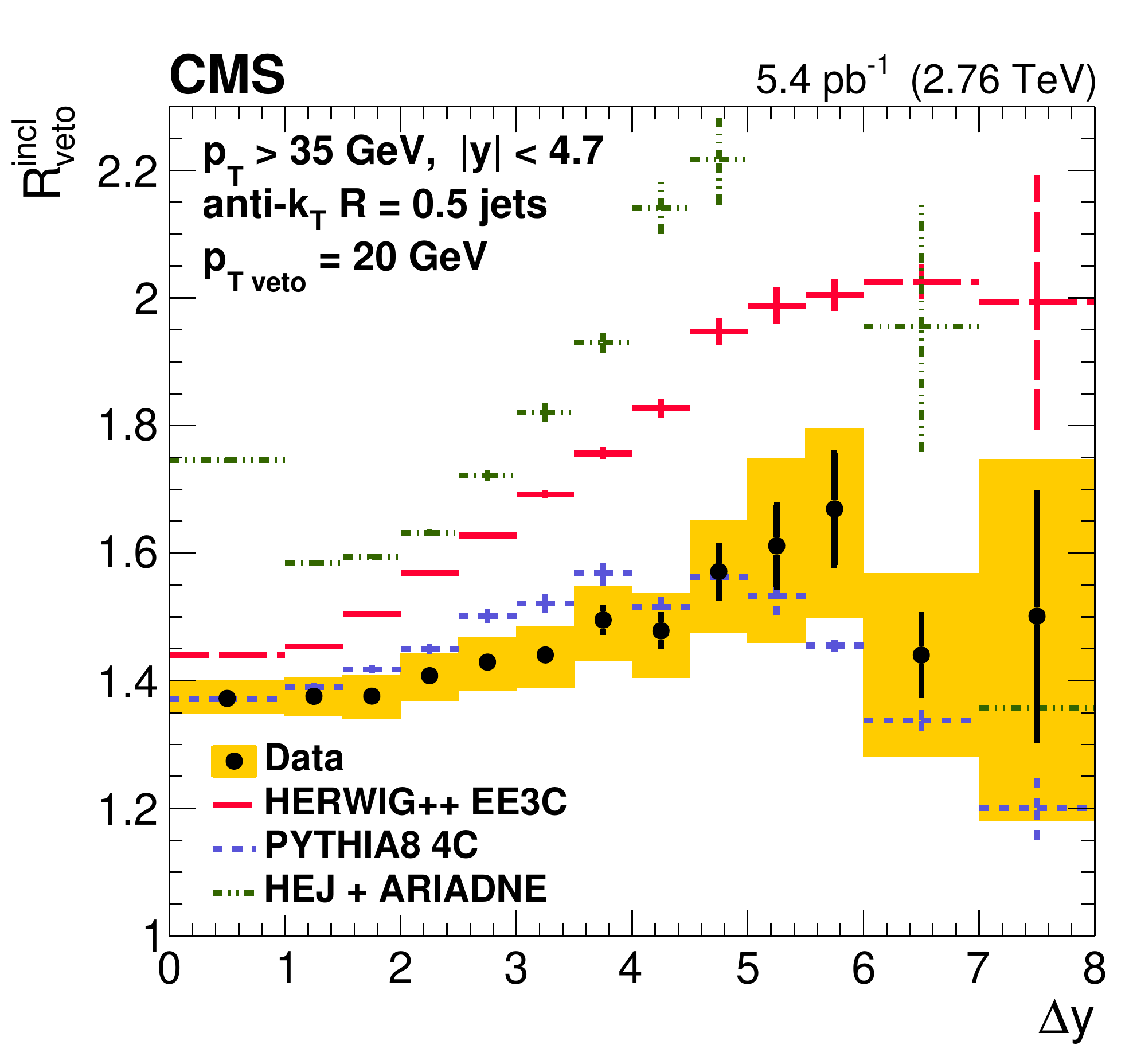}
    \includegraphics[width=0.47\textwidth]{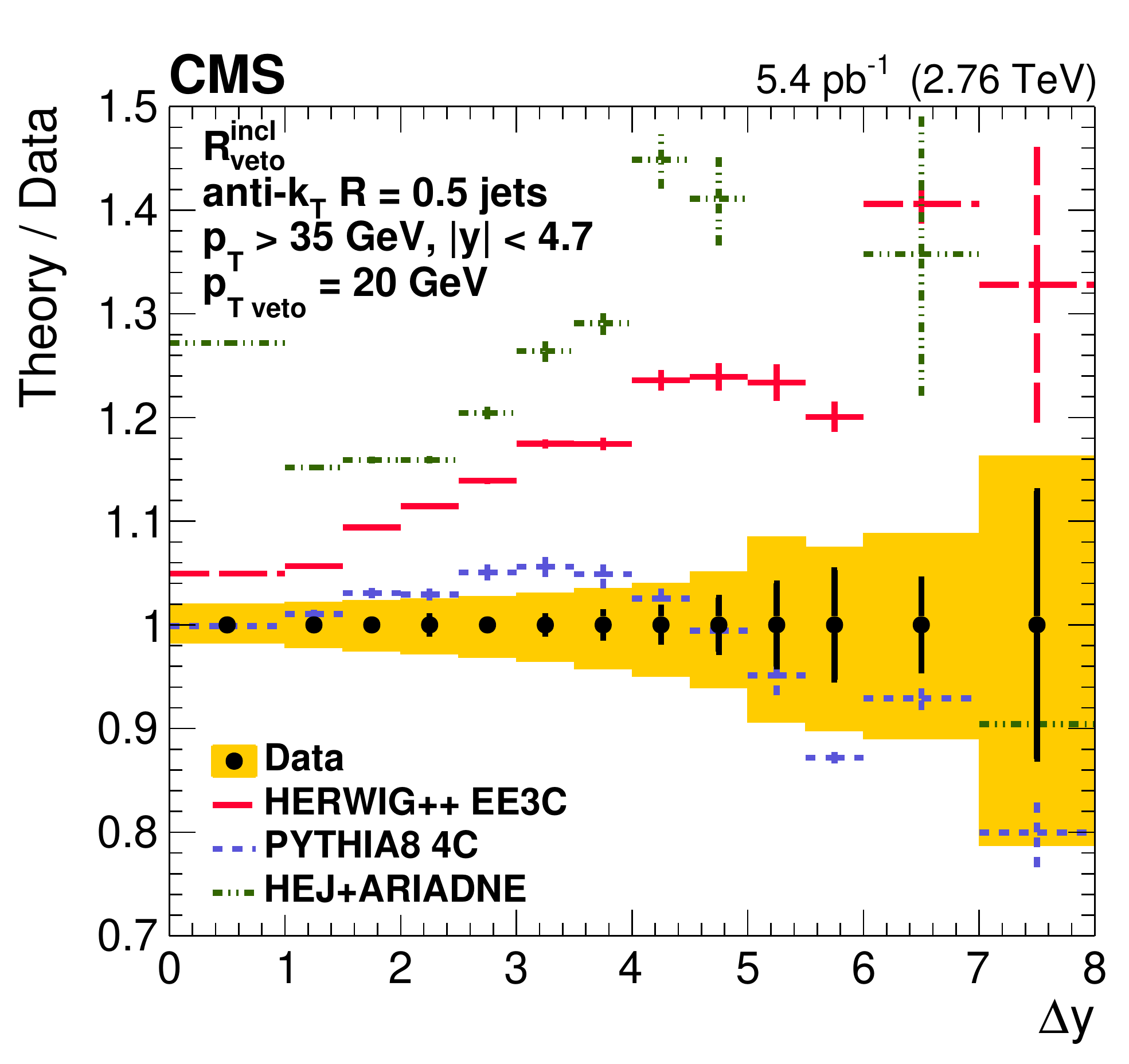}
    \includegraphics[width=0.47\textwidth]{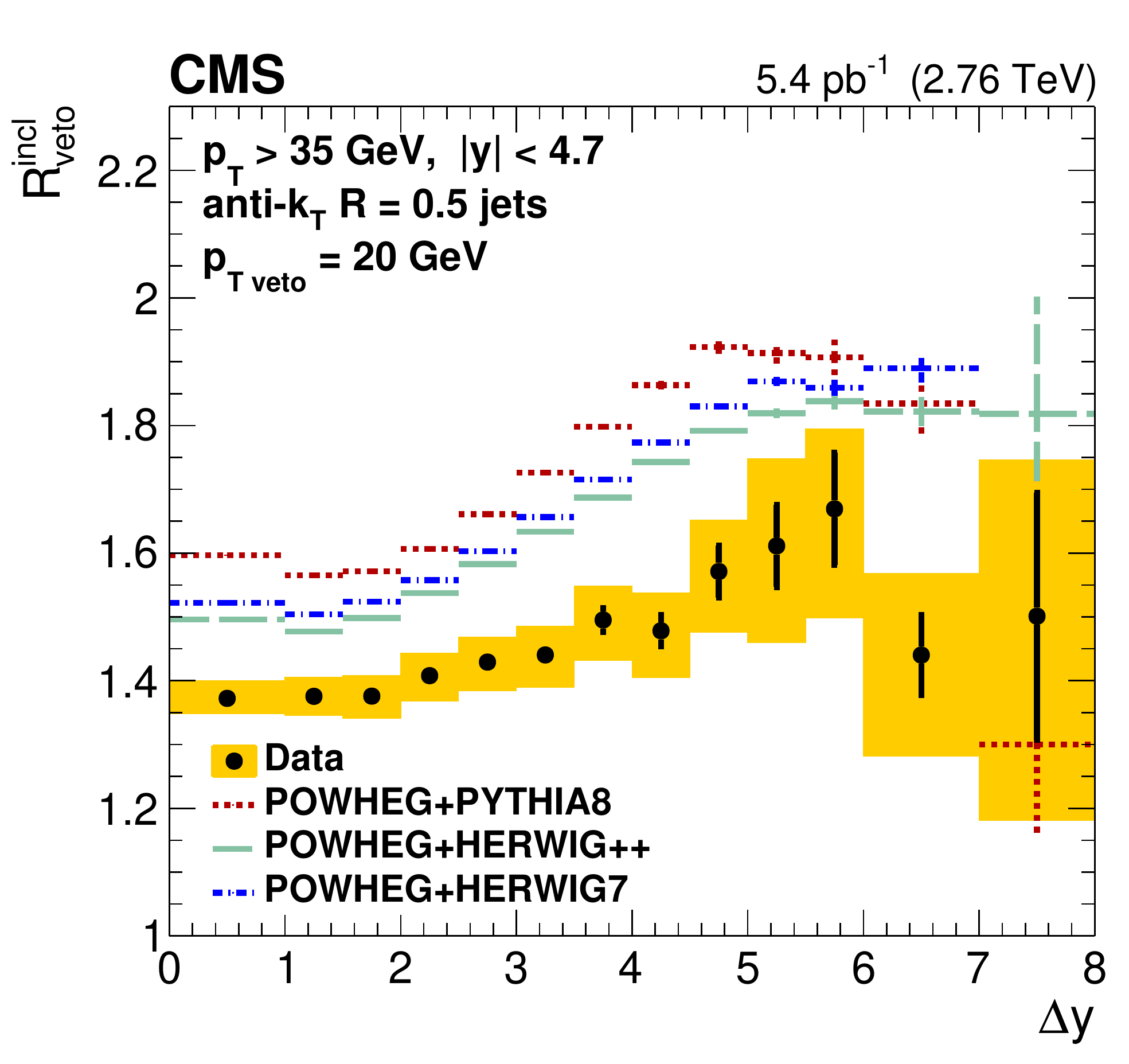}
    \includegraphics[width=0.47\textwidth]{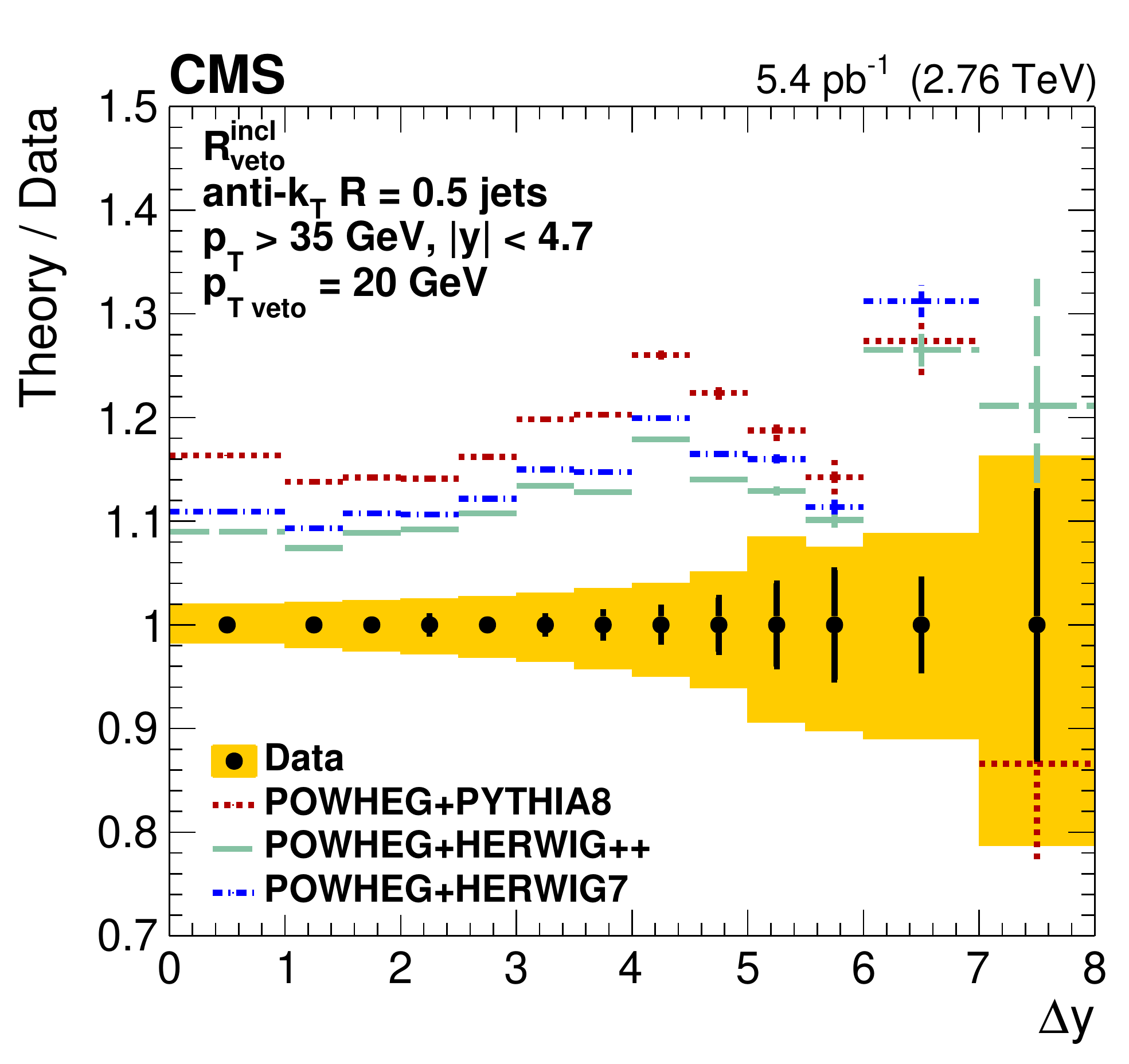}
    \caption{Ratio \Rinclveto of the cross sections for inclusive to ``exclusive" with veto dijet production. The upper and lower rows present the comparison with different MC models. The plots on the left present the ratio \Rinclveto and those on the right the ratio of theory to data. The systematic uncertainties are indicated by the shaded bands and the statistical uncertainties by the vertical bars.}
    \label{fig:k_factor_incl_veto_ak5}
\end{figure}

The ratio of the cross section for inclusive dijet production to that 
for ``exclusive" with veto dijet production, $\Rinclveto$, is presented in 
Fig.~\ref{fig:k_factor_incl_veto_ak5}. The event generators \HERWIGpp and \HEJARIADNE overestimate the measurement. \PYTHIAeight is in agreement with the data for $\Dy < 1.5$ and for $\Dy > 4$, but overestimates 
them for $2 < \Dy < 4$. If \POWHEG
is used to add the NLO corrections to the LL DGLAP-based prediction of 
\HERWIG, the description of the data becomes worse for $\Dy < 2$ and 
improves at large rapidity intervals. In the case of \POWHEGPYTHIAeight, adding the NLO corrections degrades the agreement significantly. In summary, none of the models considered reproduces all aspects of the data. Although \PYTHIAeight with tune 4C offers the best description, it still fails to model the shape of the \Dy dependence.

\subsection{Ratios \texorpdfstring{\RMN}{RMN} and \texorpdfstring{\RMNveto}{RMNveto}}
\begin{figure}[hbt!]
  \centering
    \includegraphics[width=0.47\textwidth]{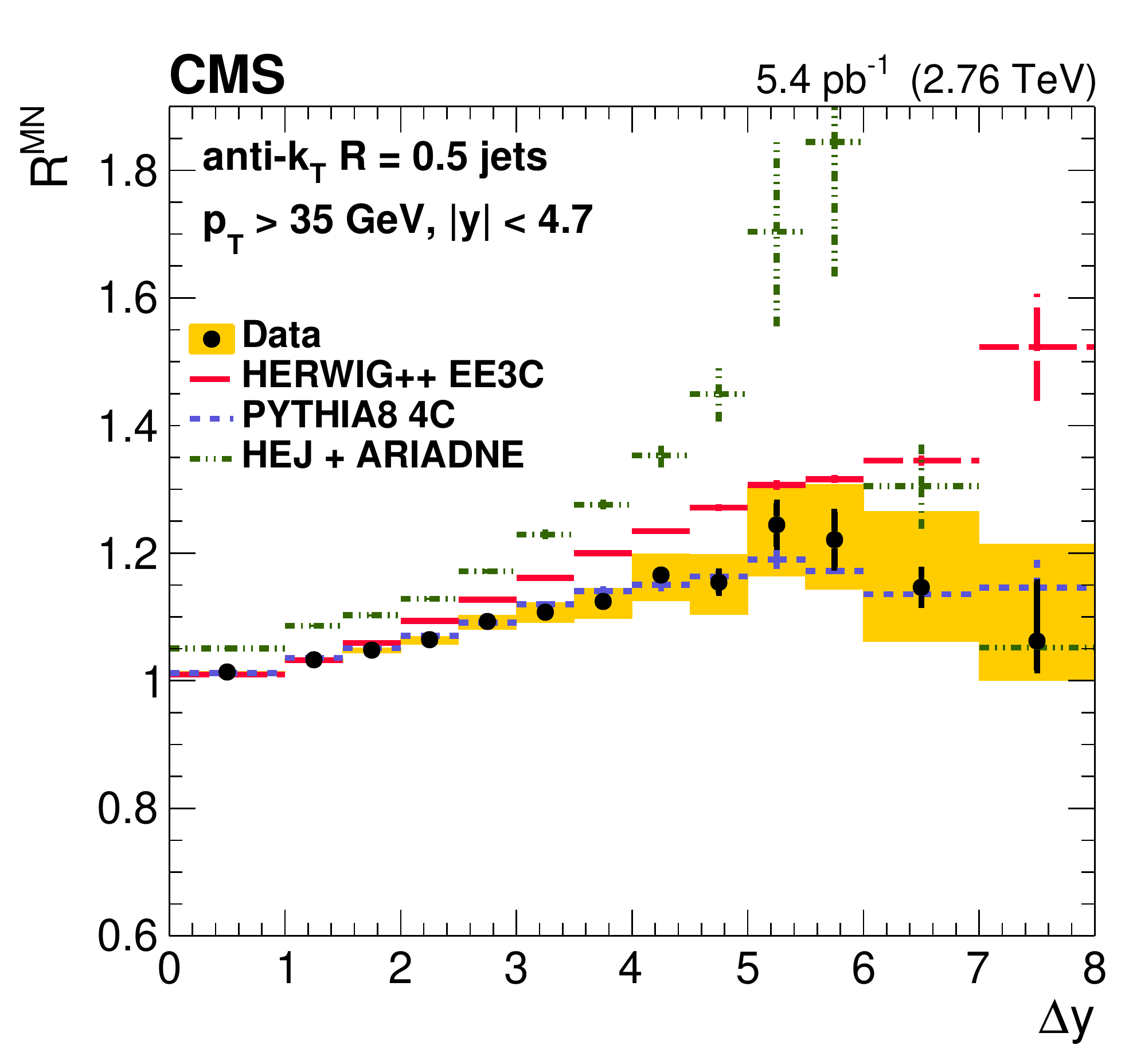}
    \includegraphics[width=0.47\textwidth]{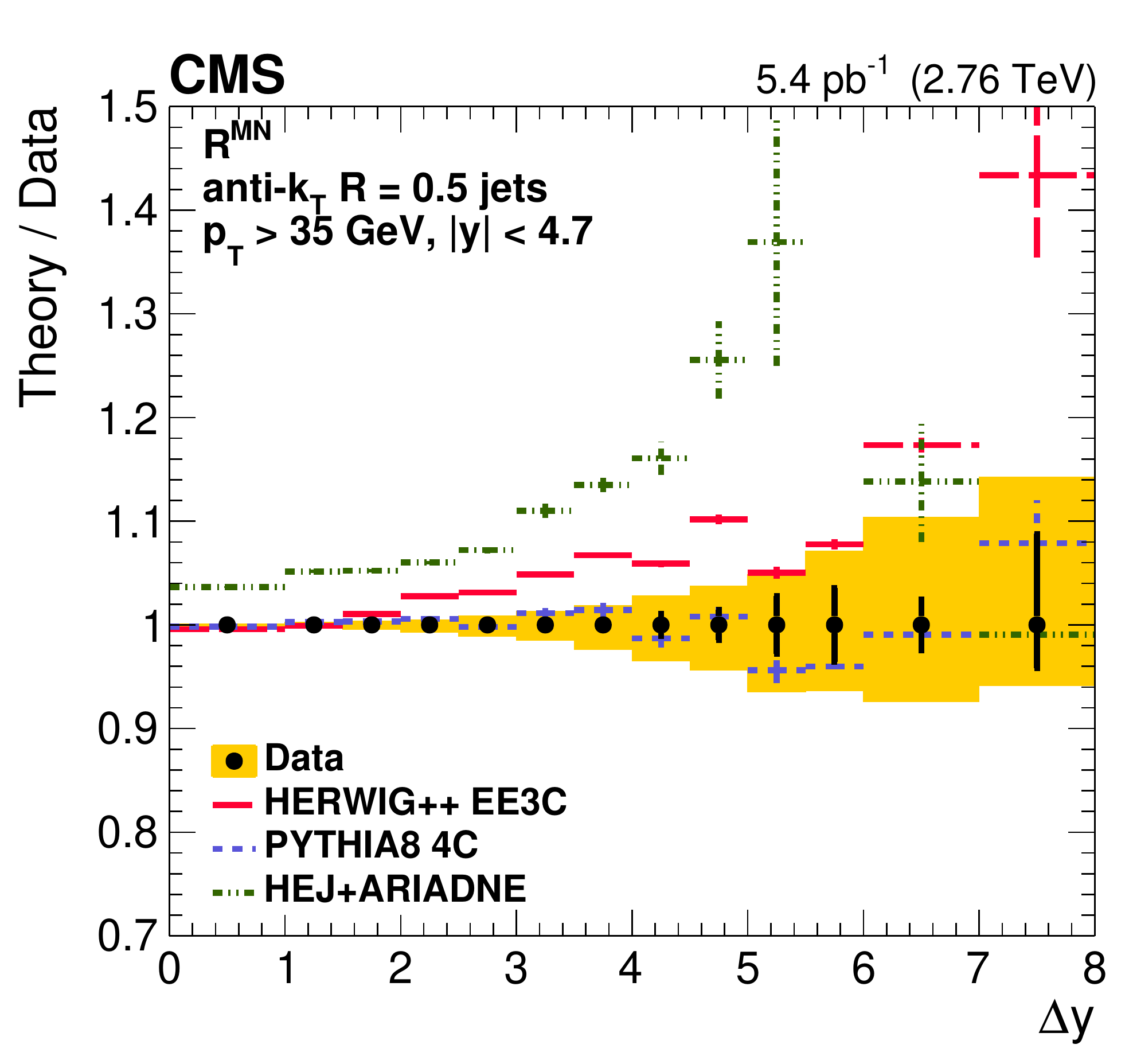}
    \includegraphics[width=0.47\textwidth]{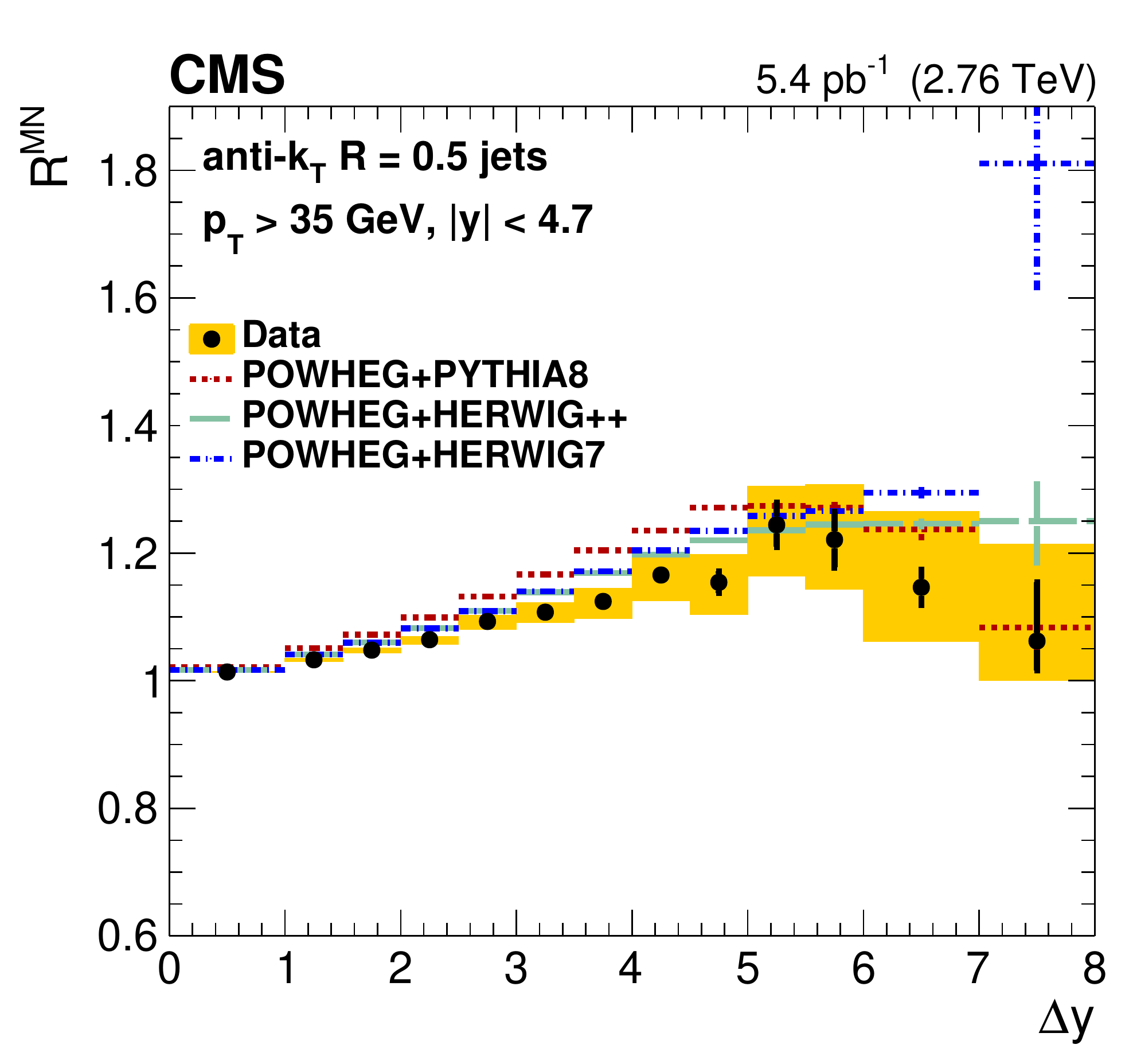}
    \includegraphics[width=0.47\textwidth]{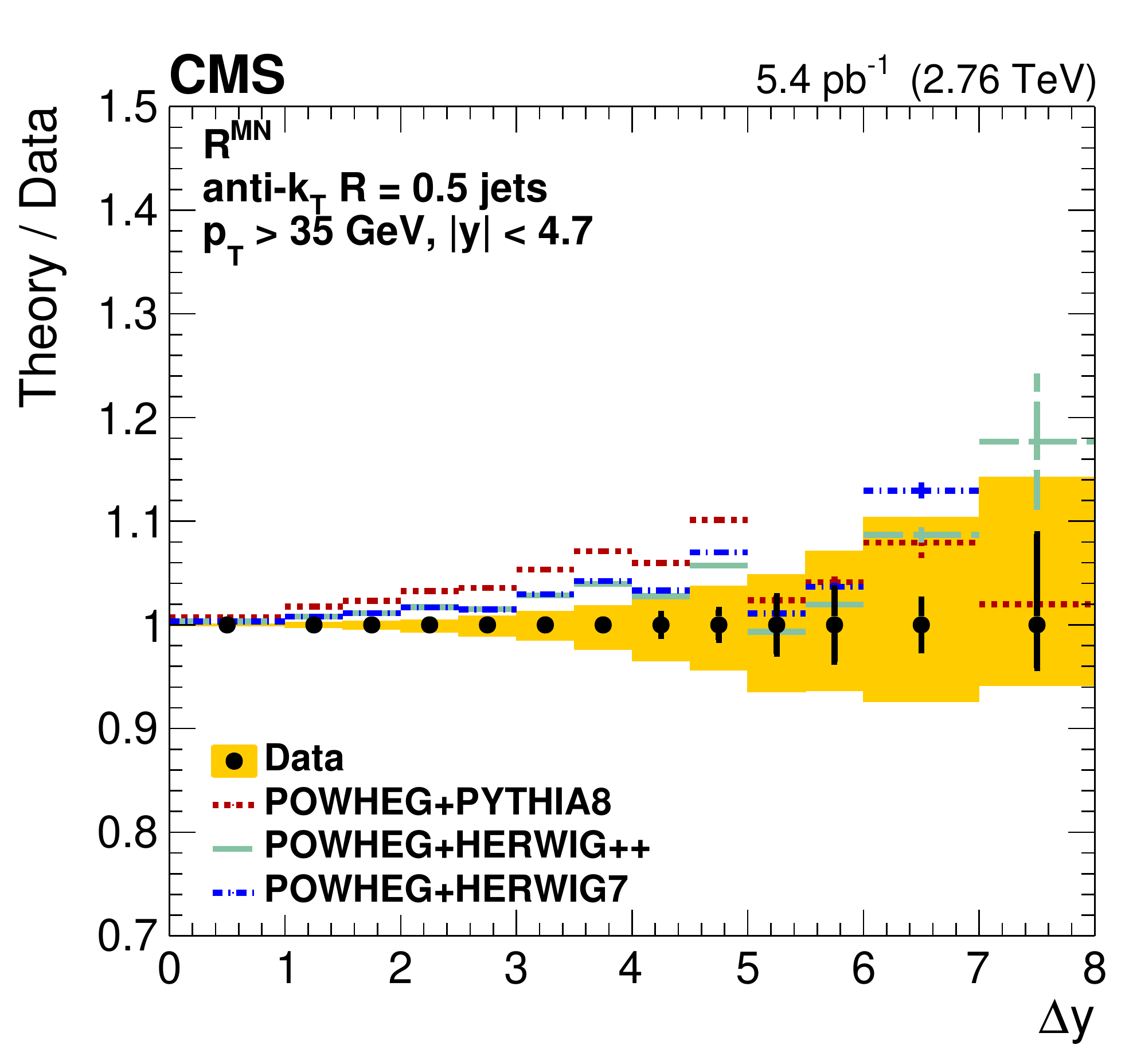}
    \caption{Ratio \RMN of the cross sections for MN to ``exclusive" dijet production. The upper and lower rows present the comparison with different MC models. The plots on the left present the ratio \RMN and those on the right the ratio of theory to data. The systematic uncertainties are indicated by the shaded bands and the statistical uncertainties by the vertical bars.}
    \label{fig:k_factor_MN_ak5}
\end{figure}

\begin{figure}[hbt!]
  \centering
    \includegraphics[width=0.47\textwidth]{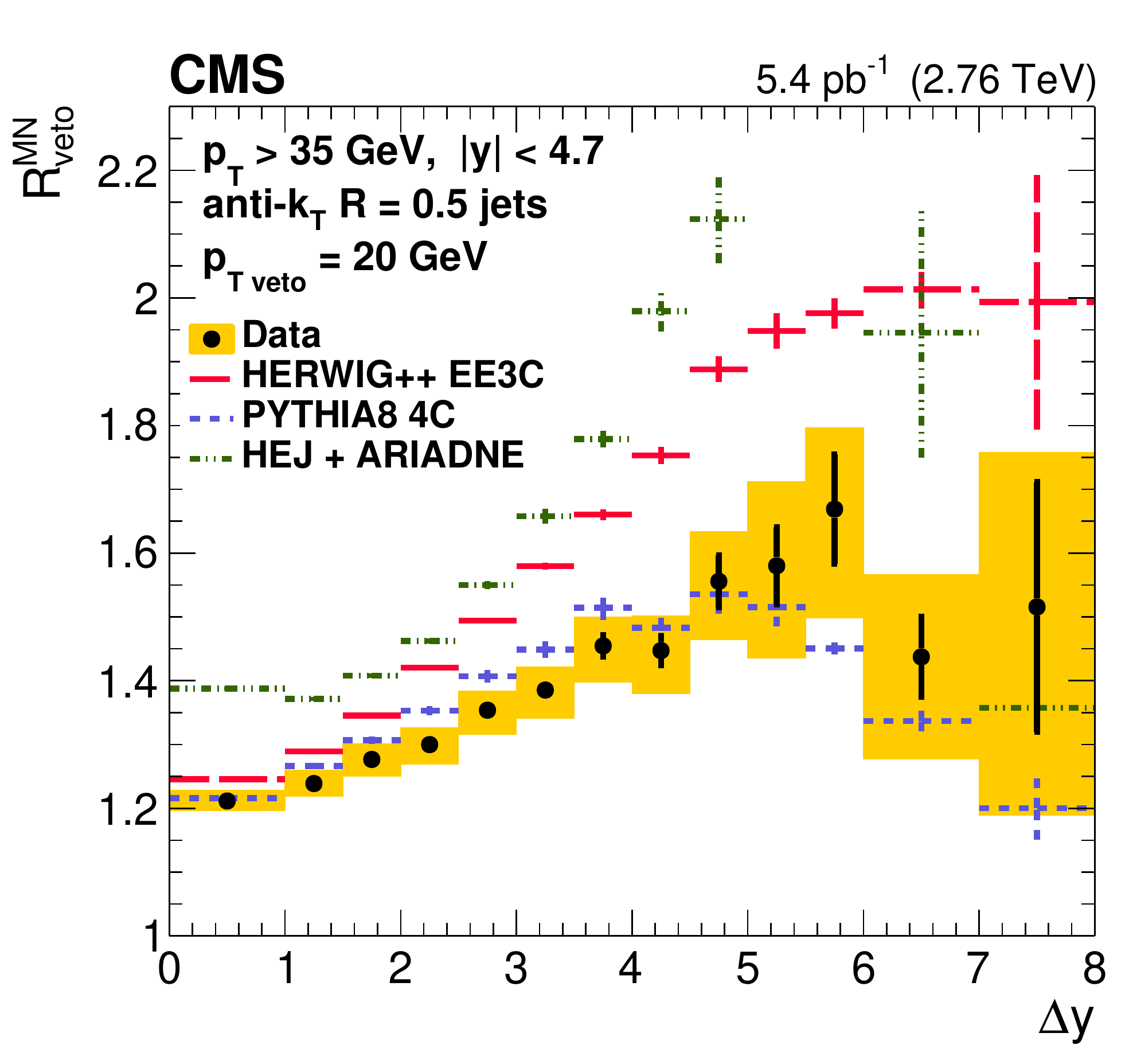}
    \includegraphics[width=0.47\textwidth]{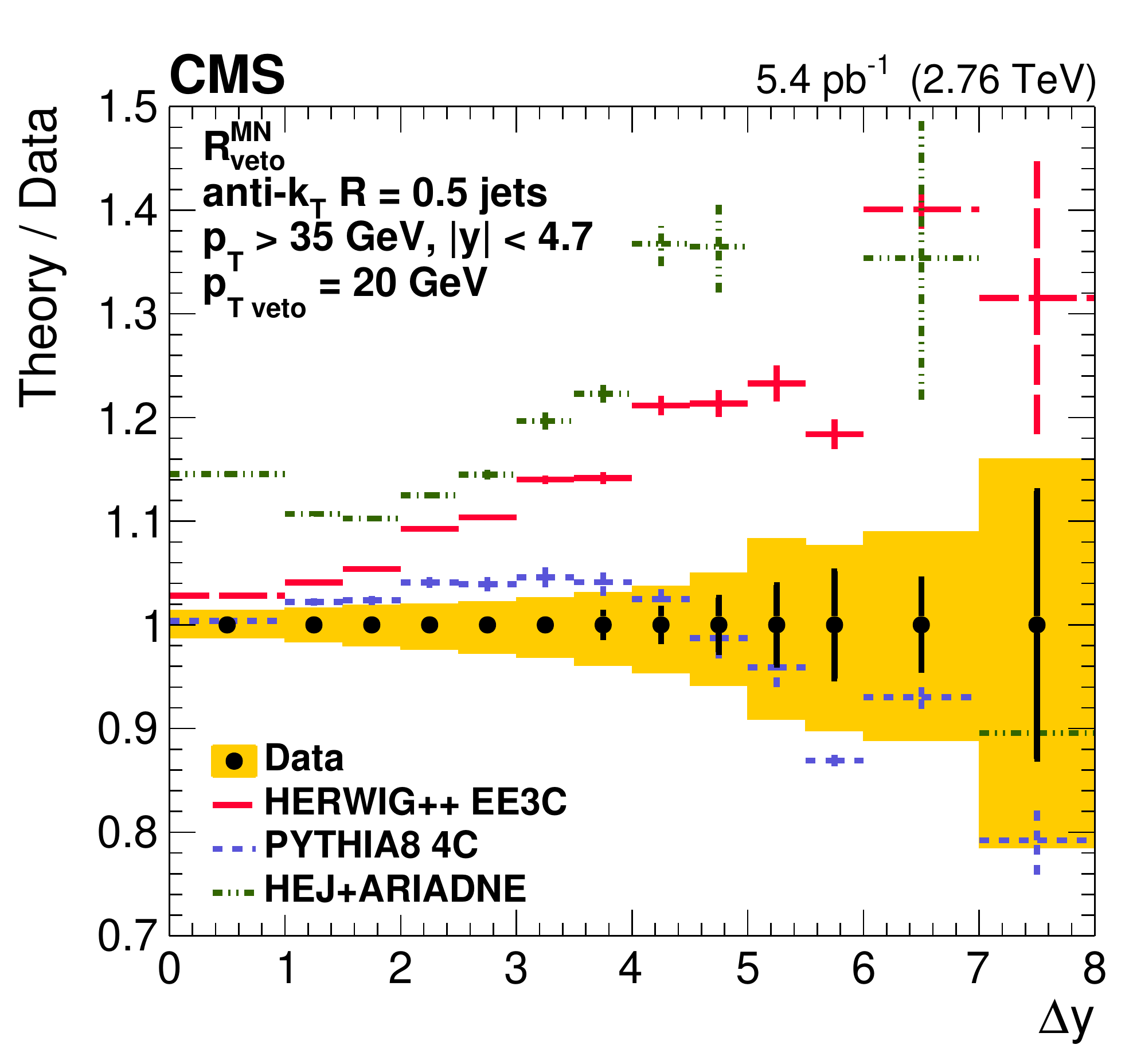}
    \includegraphics[width=0.47\textwidth]{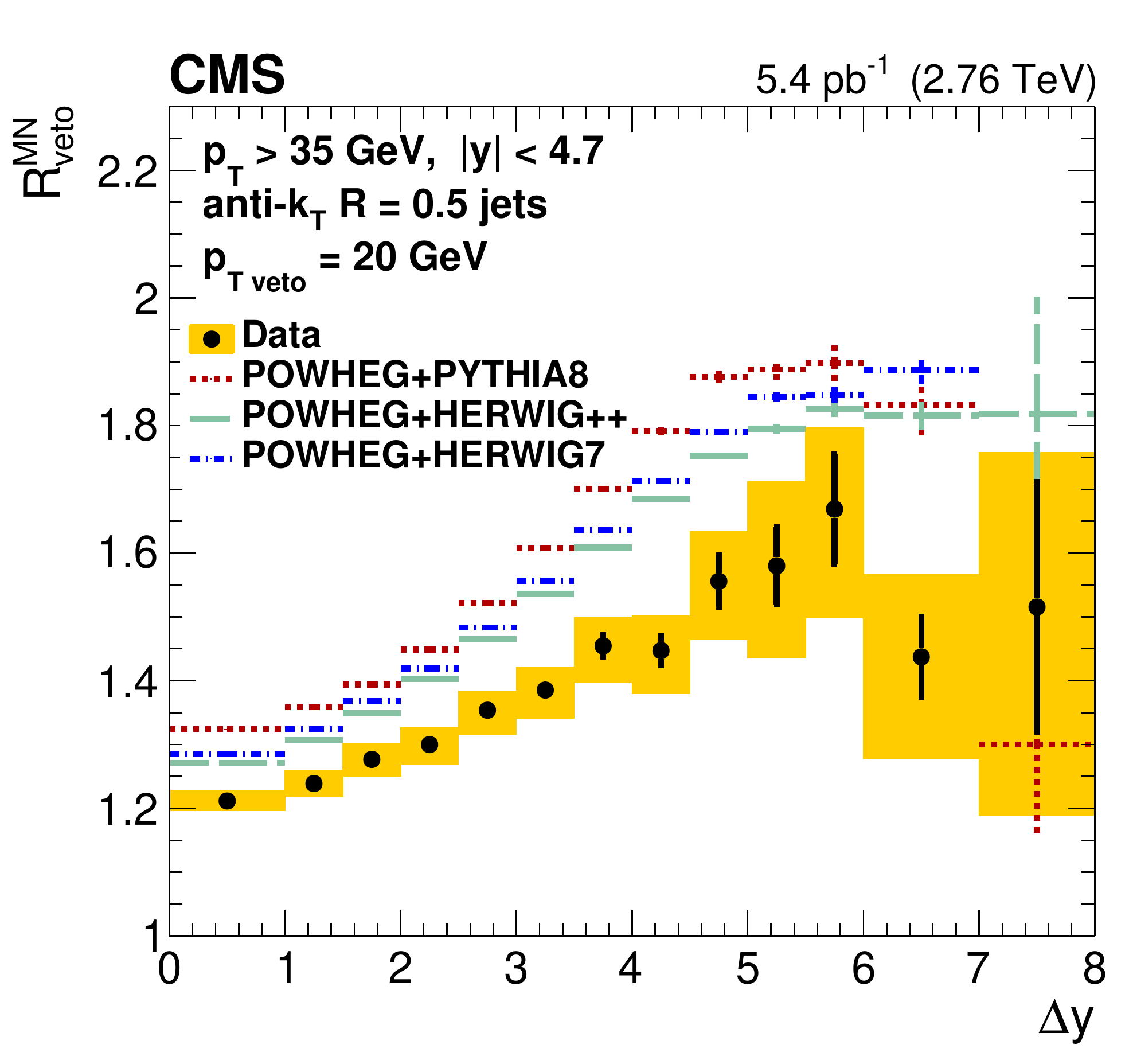}
    \includegraphics[width=0.47\textwidth]{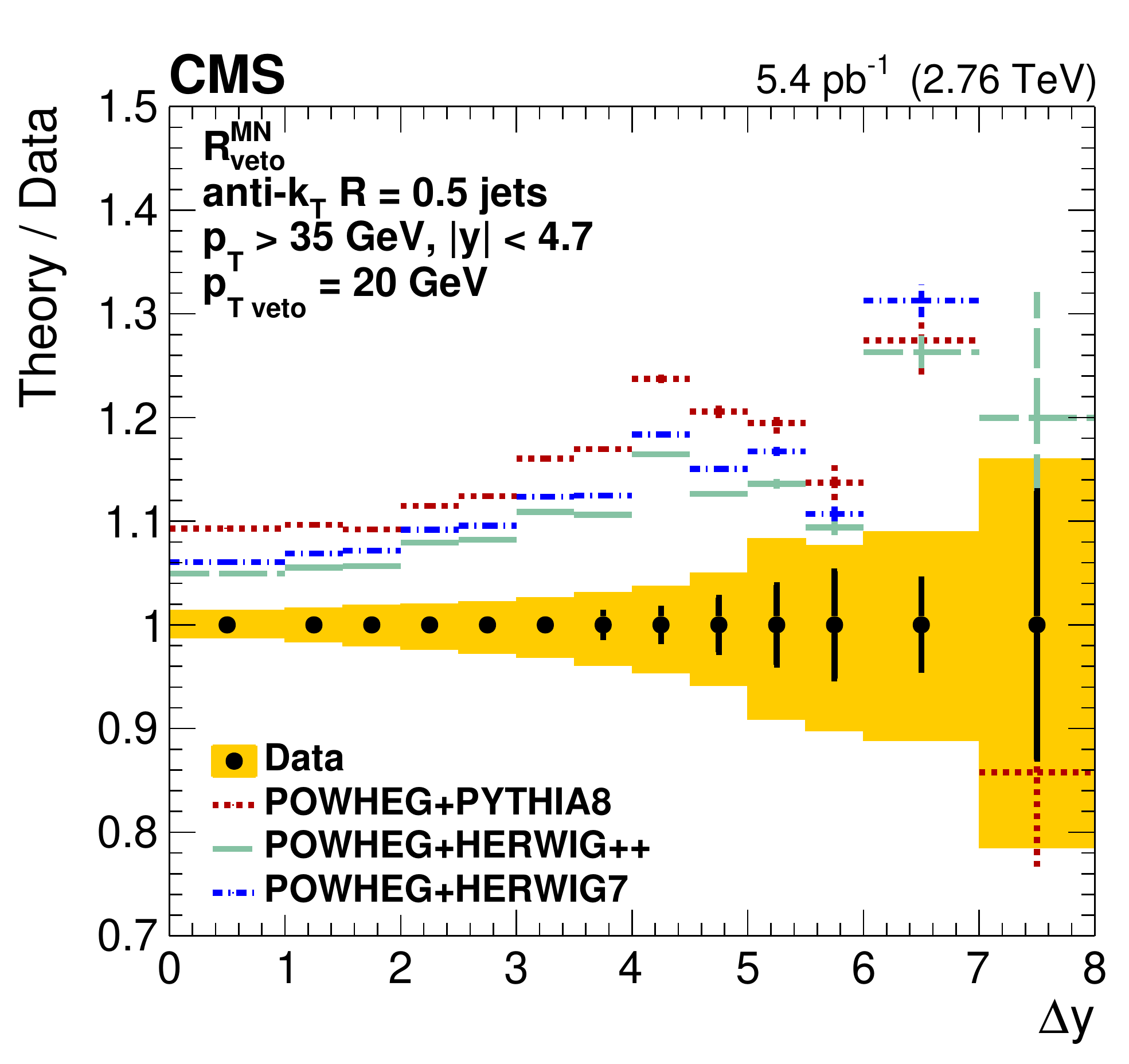}
    \caption{Ratio \RMNveto of the cross sections for MN to ``exclusive" with veto dijet production. The upper and lower rows present the comparison with different MC models. The plots on the left present the ratio \RMNveto and those on the right the ratio of theory to data. The systematic uncertainties are indicated by the shaded bands and the statistical uncertainties by the vertical bars.}
    \label{fig:k_factor_MN_veto_ak5}
\end{figure}

The ratios of the cross section for MN dijet production and those for \sgexcl as well as \sgveto, \RMN, and \RMNveto  are presented in Figs.~\ref{fig:k_factor_MN_ak5} and \ref{fig:k_factor_MN_veto_ak5}. By definition the MN ratio \RMN begins from unity at $\Dy = 0$, because there is no phase space for additional emission between the jets of the MN dijet. The MN ratio with veto \RMNveto is higher than unity at $\Dy = 0$, because of the veto condition. At large \Dy both \RMN, and \RMNveto approach \Rincl and \Rinclveto, respectively, because there is insufficient energy to produce many jets with transverse momentum above 35\GeV within the rapidity interval of the MN dijet. The results of the comparison of the MN dijet ratios with simulations are similar to those just discussed for \Rincl and \Rinclveto.

\subsection{Comparison of \texorpdfstring{\Rincl}{Rincl} and \texorpdfstring{\RMN}{RMN} at \texorpdfstring{$\sqrt{s} = 2.76$}{sqrt(s) = 2.76} and \texorpdfstring{$7\TeV$}{7 TeV}}

\begin{figure}[hbt!]
  \centering
    \includegraphics[width=0.47\textwidth]{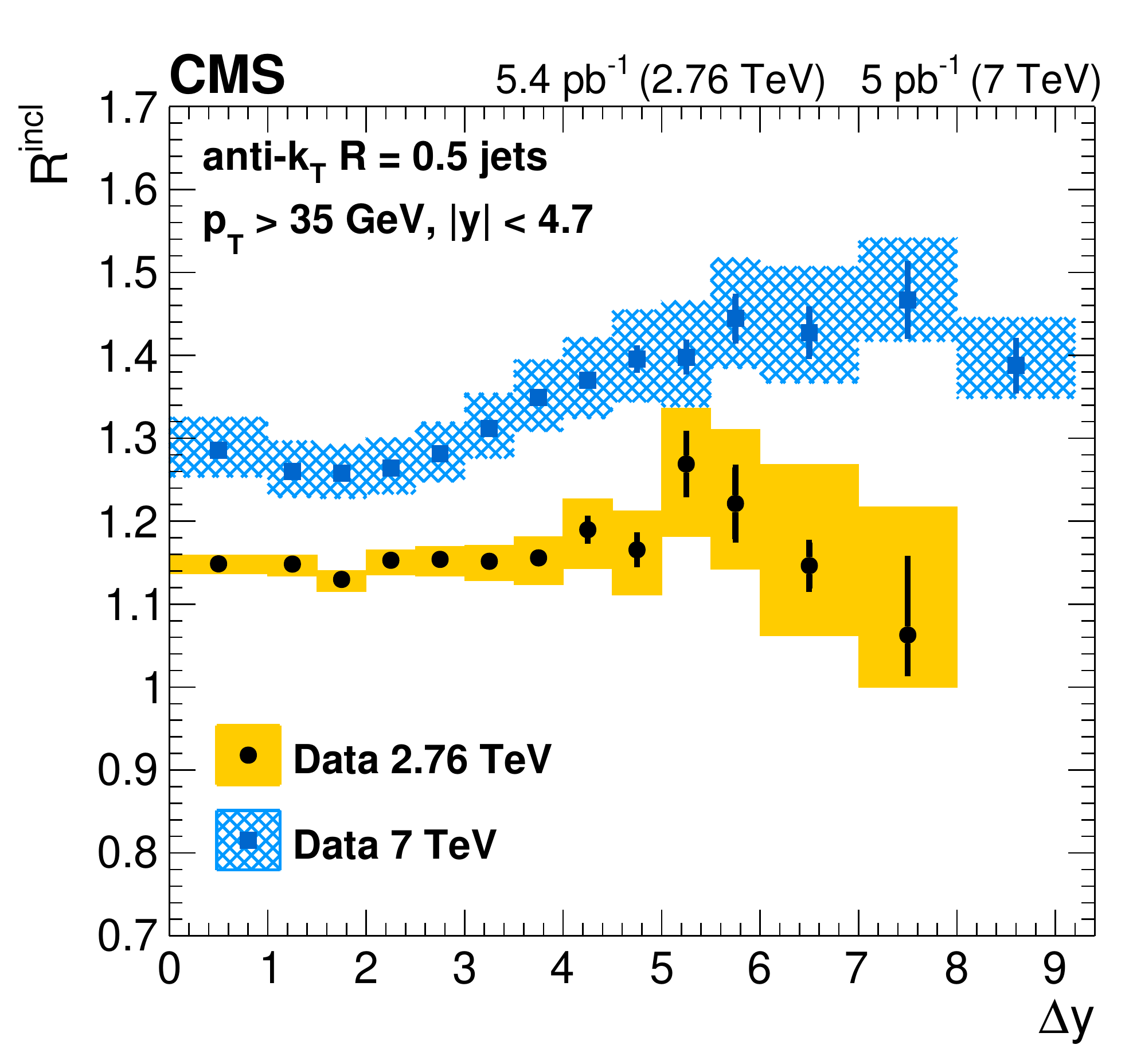}
    \includegraphics[width=0.47\textwidth]{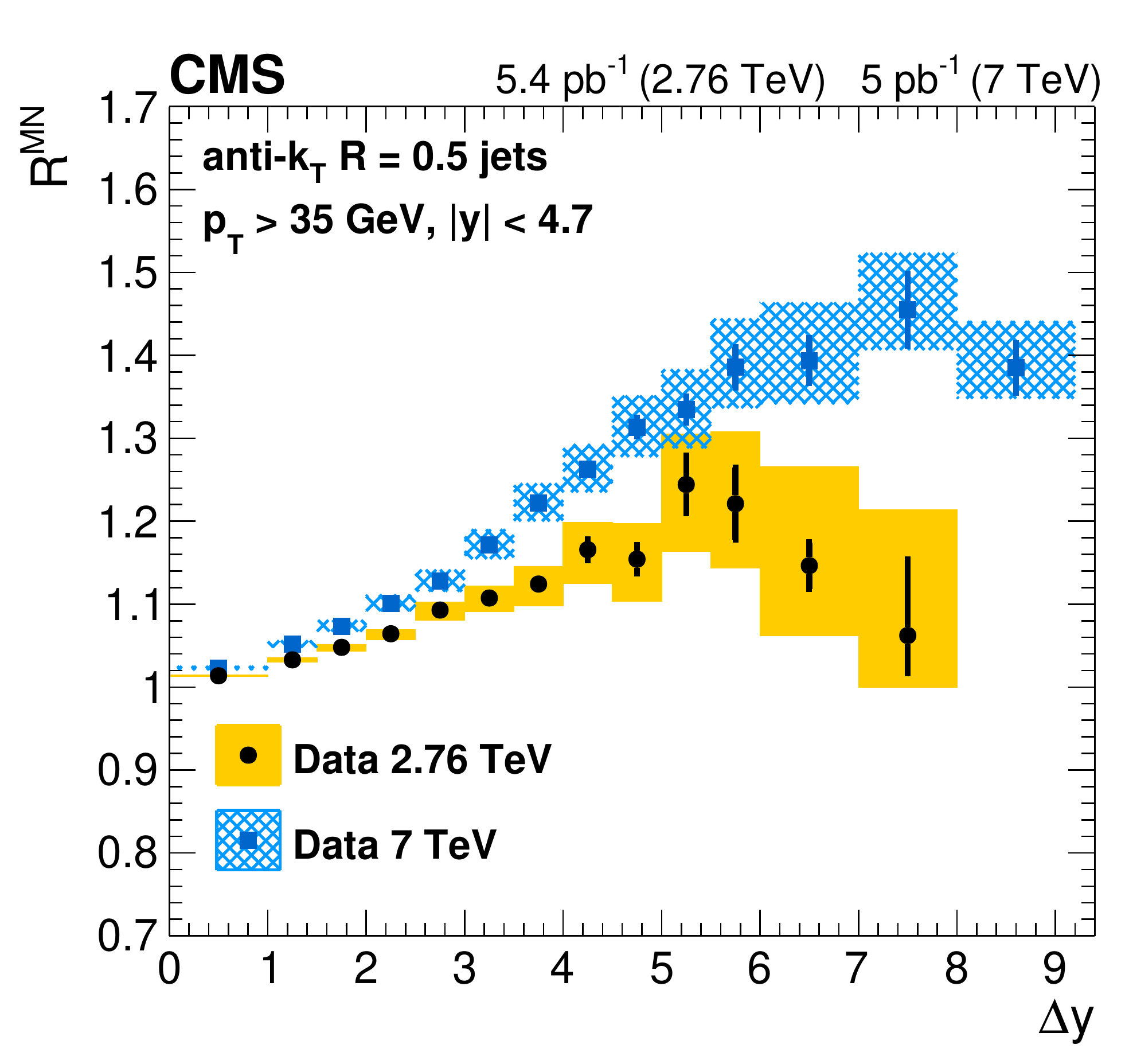}
    \caption{Ratios \Rincl (\cmsLeft) and \RMN (\cmsRight) of the cross sections for inclusive (\cmsLeft) and MN (\cmsRight) and ``exclusive" dijet production, measured at 2.76\TeV and 7\TeV~\cite{Chatrchyan:2012pb} collision energies. The systematic uncertainties are indicated by the shaded bands and the statistical uncertainties by the vertical bars.}
    \label{fig:k_factor_7TeV_comparison_ak5}
\end{figure}

The comparison of the ratios \Rincl and \RMN measured at different 
energies, $\sqrt{s} = 7\TeV$~\cite{Chatrchyan:2012pb} and 2.76\TeV, is presented in Fig.~\ref{fig:k_factor_7TeV_comparison_ak5}.
The ratios rise 
faster with \Dy at higher energy, which may reflect both 
the increasing available phase space and BFKL dynamics. Large \Dy can be more easily reached at higher energy. The qualitative features of the ratios can be understood as follows. The ratios rise with \Dy because of the increasing phase space volume for hard parton radiation. At very large \Dy, the ratios decrease because of
the kinematic limitations on the production of events with more than two
jets, each with $\pt > 35\GeV$ (for $\pt = 35 \GeV$: $\Delta \ymax=8.7$ and $10.6$ at 2.76\TeV and 7\TeV, respectively). These trends are qualitatively reproduced by the MC predictions shown in Figs.~\ref{fig:k_factor_incl_ak5}-\ref{fig:k_factor_MN_veto_ak5} as well as the corresponding MC predictions in Ref.~\cite{Chatrchyan:2012pb}. No events are observed for $\Dy > 8$ at 2.76\TeV because this region is too close to the edge of the available phase space. The comparison of the measurement with the MC models in Ref.~\cite{Chatrchyan:2012pb} shows that \PYTHIAeight gives the best
description of \Rincl and \RMN whereas
\HERWIGpp and \HEJARIADNE significantly overestimate the rise of the
ratios with \Dy, which is also consistent with the results from this analysis.

\subsection{Overall results at \texorpdfstring{$\sqrt{s} = 2.76\TeV$}{sqrt(s) = 2.76 TeV} and discussion}

In general, the measured cross sections decrease with \Dy between the jets faster than all LO DGLAP-based MC predictions. 
\PYTHIAeight overestimates the data, whereas \HERWIGpp underestimates them at small \Dy and overestimates them at large \Dy. The inclusion of the NLO corrections via \POWHEG for the LO DGLAP-based models improves the description of the inclusive (MN) cross sections at 
small \Dy only. The LL 
BFKL-based \HEJARIADNE generator underestimates the results.

The ratios \RMN and \Rincl are well described only by the LO DGLAP-based generator \PYTHIAeight. The \HERWIGpp generator calculation in the same approximation overestimates these ratios. This emphasizes the importance of corrections beyond the main approximation, such as color coherence, which can produce \Dy dependence. However, it is the BFKL resummation that consistently accounts for the principal contributions at large \Dy. The 
ratios \RMNveto and \Rinclveto cannot be perfectly reproduced by any LO DGLAP-based generator in the interval 
 $2 < \Dy < 5$. This may be an indication of BFKL effects, although the LL BFKL-based generator does not describe the data well either. The LL BFKL-based generator overestimates significantly the rise of the ratios. This can be understood because the LL BFKL calculation overestimates the intercept of the pomeron, which leads to a stronger rise of cross sections with \Dy. The analysis for the NLL BFKL contributions, with a lower pomeron intercept, is important ~\cite{Brodsky:1998kn}.   
Since an analytic NLL BFKL evolution
 calculation~\cite{Ducloue:2013bva} provides a good description of the decorrelations of MN dijets~\cite{Khachatryan:2016udy},
 we also need an analogous prediction 
 (based on the NLL BFKL evolution) to draw a more firm conclusion from our measurement.

In conclusion, none of the LO or NLO DGLAP-based generators can provide a reasonable description of the measured cross sections and their ratios simultaneously. The dijet ratio with extra jet veto cannot be described by DGLAP-based generators either at LO or at NLO. 
For a useful comparison of the present data with the BFKL approach, 
calculations at NLL are needed.

\section{Summary}

A study of dijet events with large rapidity separation \Dy between the jets 
has been performed using proton-proton collision data at $\sqrt{s} = 2.76\TeV$, collected by the CMS experiment in 2013 with an integrated luminosity of 5.4\pbinv. The cross sections for Mueller--Navelet and inclusive dijet event 
production, as well as their ratios to ``exclusive", and ``exclusive" with veto, dijet event production, 
are measured up to $\Dy \leq 8.0$ between the jets. 

None of the Dokshitzer--Gribov--Lipatov--Altarelli--Parisi (DGLAP)-based Monte Carlo generators using leading-order (LO) or next-to-LO (NLO) calculations can provide a complete description of all measured cross sections and their ratios. The dijet ratio with extra jet veto cannot be described by the LO or NLO DGLAP-based generators.
To compare the present results with the Balitsky--Fadin--Kuraev--Lipatov approach, calculations at the next-to-leading-logarithm level are needed.

The present results at $\sqrt{s} = 2.76\TeV$ can be used along with data at higher energies to reveal possible effects beyond the DGLAP approach to make more definite conclusions. Tabulated results are provided in the HEPData record for this analysis~\cite{hepdata}.

\begin{acknowledgments}
  We congratulate our colleagues in the CERN accelerator departments for the excellent performance of the LHC and thank the technical and administrative staffs at CERN and at other CMS institutes for their contributions to the success of the CMS effort. In addition, we gratefully acknowledge the computing  centers and personnel of the Worldwide LHC Computing Grid and other  centers for delivering so effectively the computing infrastructure essential to our analyses. Finally, we acknowledge the enduring support for the construction and operation of the LHC, the CMS detector, and the supporting computing infrastructure provided by the following funding agencies: BMBWF and FWF (Austria); FNRS and FWO (Belgium); CNPq, CAPES, FAPERJ, FAPERGS, and FAPESP (Brazil); MES and BNSF (Bulgaria); CERN; CAS, MoST, and NSFC (China); MINCIENCIAS (Colombia); MSES and CSF (Croatia); RIF (Cyprus); SENESCYT (Ecuador); MoER, ERC PUT and ERDF (Estonia); Academy of Finland, MEC, and HIP (Finland); CEA and CNRS/IN2P3 (France); BMBF, DFG, and HGF (Germany); GSRI (Greece); NKFIA (Hungary); DAE and DST (India); IPM (Iran); SFI (Ireland); INFN (Italy); MSIP and NRF (Republic of Korea); MES (Latvia); LAS (Lithuania); MOE and UM (Malaysia); BUAP, CINVESTAV, CONACYT, LNS, SEP, and UASLP-FAI (Mexico); MOS (Montenegro); MBIE (New Zealand); PAEC (Pakistan); MSHE and NSC (Poland); FCT (Portugal); JINR (Dubna); MON, RosAtom, RAS, RFBR, and NRC KI (Russia); MESTD (Serbia); SEIDI, CPAN, PCTI, and FEDER (Spain); MOSTR (Sri Lanka); Swiss Funding Agencies (Switzerland); MST (Taipei); ThEPCenter, IPST, STAR, and NSTDA (Thailand); TUBITAK and TAEK (Turkey); NASU (Ukraine); STFC (United Kingdom); DOE and NSF (USA).

  \hyphenation{Rachada-pisek} Individuals have received support from the Marie-Curie program and the European Research Council and Horizon 2020 Grant, contract Nos.\ 675440, 724704, 752730, 758316, 765710, 824093, 884104, and COST Action CA16108 (European Union); the Leventis Foundation; the Alfred P.\ Sloan Foundation; the Alexander von Humboldt Foundation; the Belgian Federal Science Policy Office; the Fonds pour la Formation \`a la Recherche dans l'Industrie et dans l'Agriculture (FRIA-Belgium); the Agentschap voor Innovatie door Wetenschap en Technologie (IWT-Belgium); the F.R.S.-FNRS and FWO (Belgium) under the ``Excellence of Science -- EOS" -- be.h project n.\ 30820817; the Beijing Municipal Science \& Technology Commission, No. Z191100007219010; the Ministry of Education, Youth and Sports (MEYS) of the Czech Republic; the Deutsche Forschungsgemeinschaft (DFG), under Germany's Excellence Strategy -- EXC 2121 ``Quantum Universe" -- 390833306, and under project number 400140256 - GRK2497; the Lend\"ulet (``Momentum") Program and the J\'anos Bolyai Research Scholarship of the Hungarian Academy of Sciences, the New National Excellence Program \'UNKP, the NKFIA research grants 123842, 123959, 124845, 124850, 125105, 128713, 128786, and 129058 (Hungary); the Council of Science and Industrial Research, India; the Latvian Council of Science; the Ministry of Science and Higher Education and the National Science Center, contracts Opus 2014/15/B/ST2/03998 and 2015/19/B/ST2/02861 (Poland); the Funda\c{c}\~ao para a Ci\^encia e a Tecnologia, grant CEECIND/01334/2018 (Portugal); the National Priorities Research Program by Qatar National Research Fund; the Ministry of Science and Higher Education, projects no. 14.W03.31.0026 and no. FSWW-2020-0008, and the Russian Foundation for Basic Research, project No.19-42-703014 (Russia); the Programa Estatal de Fomento de la Investigaci{\'o}n Cient{\'i}fica y T{\'e}cnica de Excelencia Mar\'{\i}a de Maeztu, grant MDM-2015-0509 and the Programa Severo Ochoa del Principado de Asturias; the Stavros Niarchos Foundation (Greece); the Rachadapisek Sompot Fund for Postdoctoral Fellowship, Chulalongkorn University and the Chulalongkorn Academic into Its 2nd Century Project Advancement Project (Thailand); the Kavli Foundation; the Nvidia Corporation; the SuperMicro Corporation; the Welch Foundation, contract C-1845; and the Weston Havens Foundation (USA).
\end{acknowledgments}

\bibliography{auto_generated}

\cleardoublepage \appendix\section{The CMS Collaboration \label{app:collab}}\begin{sloppypar}\hyphenpenalty=5000\widowpenalty=500\clubpenalty=5000\cmsinstitute{Yerevan~Physics~Institute, Yerevan, Armenia}
A.~Tumasyan
\cmsinstitute{Institut~f\"{u}r~Hochenergiephysik, Vienna, Austria}
W.~Adam\cmsorcid{0000-0001-9099-4341}, J.W.~Andrejkovic, T.~Bergauer\cmsorcid{0000-0002-5786-0293}, S.~Chatterjee\cmsorcid{0000-0003-2660-0349}, M.~Dragicevic\cmsorcid{0000-0003-1967-6783}, A.~Escalante~Del~Valle\cmsorcid{0000-0002-9702-6359}, R.~Fr\"{u}hwirth\cmsAuthorMark{1}, M.~Jeitler\cmsAuthorMark{1}\cmsorcid{0000-0002-5141-9560}, N.~Krammer, L.~Lechner\cmsorcid{0000-0002-3065-1141}, D.~Liko, I.~Mikulec, F.M.~Pitters, J.~Schieck\cmsAuthorMark{1}\cmsorcid{0000-0002-1058-8093}, R.~Sch\"{o}fbeck\cmsorcid{0000-0002-2332-8784}, M.~Spanring\cmsorcid{0000-0001-6328-7887}, S.~Templ\cmsorcid{0000-0003-3137-5692}, W.~Waltenberger\cmsorcid{0000-0002-6215-7228}, C.-E.~Wulz\cmsAuthorMark{1}\cmsorcid{0000-0001-9226-5812}
% ...
\vskip\cmsinstskip
% ...
1:  Also at TU~Wien, Wien, Austria\\
% ...
\end{sloppypar}
\end{document}